\documentclass[a4paper,11pt]{article}
\pdfoutput=1 

\usepackage{jheppub}

\usepackage[T1]{fontenc}

\usepackage{amsmath,amssymb,amsfonts, physics}
\usepackage{mathrsfs}
\usepackage{bbm}
\usepackage{slashed}
\usepackage{graphicx}
\usepackage{subcaption}
\usepackage{mwe}
\usepackage{verbatim}
\usepackage[T1]{fontenc}
\usepackage[utf8]{inputenc}
\usepackage[dvipsnames]{xcolor}
\usepackage[normalem]{ulem}
\usepackage{svg}

\usepackage{tikz}
\usepackage{tikz-3dplot}
\usetikzlibrary{shapes.geometric, 3d, calc, decorations.pathmorphing}
\tdplotsetmaincoords{75}{110}

\usepackage[backend=bibtex,style=numeric-comp, sorting=none]{biblatex}
\bibliography{references}

\definecolor{pink}{RGB}{255,192,203}

\def\be#1\ee{\begin{align}#1\end{align}} 
\newcommand{\rmd}{\mathrm{d}} 
\newcommand{\qu}[1]{``{#1}''} 
\newcommand{\GN}{G_\text{N}}       
\newcommand{\GC}{G_{\text{C}^2}}   
\newcommand{\gc}{g_{\text{C}^2}}   
\newcommand{\gci}{g_{\text{C}^2,\text{IR}}}   
\newcommand{\longbar}[1]{\,\overline{\!{#1}}} 

\title{\boldmath Charting GLOBs in Asymptotically Safe Gravity}

\author[a,b]{Francesco Del Porro,}
\author[a,b]{Jonas Pfeiffer,}
\author[a,b]{Alessia Platania,}
\author[c,d,e]{Samuele Silveravalle}
\affiliation[a]{Center of Gravity, Niels Bohr Institute, Blegdamsvej 17, DK-2100 Copenhagen \O, DENMARK}
\affiliation[b]{Niels Bohr International Academy, Niels Bohr Institute, Blegdamsvej 17, DK-2100 Copenhagen \O, DENMARK}
\affiliation[c]{SISSA - International School for Advanced Studies, Via Bonomea 265, 34136 Trieste, Italy}
\affiliation[d]{INFN, Sezione di Trieste, Via Valerio 2, 34127 Trieste, Italy}
\affiliation[e]{IFPU - Institute for Fundamental Physics of the Universe, Via Beirut 2, 34151 Trieste, Italy}

\emailAdd{francesco.del.porro@nbi.ku.dk}
\emailAdd{mwz422@alumni.ku.dk}
\emailAdd{alessia.platania@nbi.ku.dk}
\emailAdd{ssilvera@sissa.it}

\abstract{Computing the gravitational effective action provides a direct route to charting the landscape of admissible black hole spacetimes and their alternatives, which we will collectively call ``gravitationally localized objects'' (GLOBs). In this work, we provide a proof of principle of this idea within the framework of asymptotically safe gravity. Focusing on the Einstein-Weyl truncation, we identify the unique ultraviolet-complete trajectory emanating from the asymptotically safe fixed point and use it to extract the Wilson coefficient of the Weyl-squared term. This allows us to chart the corresponding GLOBs in a ``phase diagram'', showing that wormholes dominate a large portion of it, whereas the classical Bachian naked singularities become disfavored. Our results illustrate how quantum gravity can constrain effective field theory and the associated set of allowed spacetimes, yielding a rich landscape of beyond-general-relativity solutions rather than a single alternative to classical black holes.}

\begin{document} 
\maketitle
\flushbottom

\newpage
\section{Introduction}\label{sect:introduction}

The quest for a consistent and predictive quantum theory of gravity remains one of the most profound open problems in modern theoretical physics. Despite its unparalleled success on macroscopic scales, Einstein's general relativity (GR) breaks down at extremely short distances, where quantum effects are expected to dominate. Black holes in particular offer an important arena to probe the interface between gravity and quantum mechanics: their thermodynamical properties are deeply connected to quantum aspects of gravity, while their global structure, including whether they are actually black holes or alternative spacetimes, is strongly tied to the classical and quantum modifications of the Einstein-Hilbert dynamics.

In the absence of a complete and universally accepted theory of quantum gravity (QG)~\cite{Bambi2024-nm,Basile:2024oms,Buoninfante:2024yth}, various models have been proposed to describe ``quantum-corrected'' black holes (see~\cite{Buoninfante:2024oxl,Carballo-Rubio:2025fnc} for recent overviews). Early efforts focused on constructing effective geometries that resolve classical singularities while preserving essential features such as the Newtonian limit and thermodynamic consistency. Seminal examples include the Dymnikova~\cite{Dymnikova:1992ux} and Hayward~\cite{Hayward:2005gi} metrics, which modify the Schwarzschild geometry by introducing scale parameters associated with the length scale of new physics. These constructions, while physically motivated, are inherently phenomenological: they are built to incorporate expected features rather than derived from an underlying microscopic dynamics.
Approaches inspired by candidate theories of QG, such as string theory (ST)~\cite{Polchinski:1998rq,Polchinski:1998rr}, loop quantum gravity (LQG)~\cite{Rovelli:1997yv,Ashtekar:2021kfp}, and asymptotically safe quantum gravity (ASQG)~\cite{Percacci:2017fkn,Reuter:2019byg}, have also been employed.

Among the different QG theories, ASQG~\cite{Knorr:2022dsx, Eichhorn:2022gku, Morris:2022btf, Martini:2022sll, Wetterich:2022ncl, Platania:2023srt, Saueressig:2023irs, Pawlowski:2023gym, Bonanno:2024xne} is arguably one of the most conservative scenarios and the focus of our work: originally proposed by Weinberg in the late 1970s~\cite{Weinberg:1980gg}, it posits that gravity can be ultraviolet (UV) complete within a quantum field theory (QFT) framework if its renormalization group (RG) flow is governed by a non-Gaussian fixed point (NGFP) at high energies,  with a finite number of relevant directions. Approaching such a fixed point, the dimensionless couplings tend to finite values, and the number of relevant directions relates to the number of free parameters describing the corresponding low-energy effective field theory (EFT). Over the past decades, mounting evidence for asymptotic safety has been obtained both via the lattice Monte Carlo simulations of dynamical triangulations~\cite{Loll:2019rdj,Dai:2023tud} and via the semi-analytical methods of the functional renormalization group (FRG)~\cite{Dupuis:2020fhh}. This nonperturbative method, based on the Wetterich equation for the effective average action~\cite{Wetterich:1992yh}, allows for a systematic exploration of RG flows in theory space, and only requires an ansatz for the action as an input. Initial studies focused on the Einstein-Hilbert truncation of the gravitational effective action~\cite{Reuter:1996cp,Souma:1999at}, as well as its coupling to matter fields. These investigations have shown that a suitable NGFP exists, even when coupling gravity with the whole field content of the Standard Model~\cite{Eichhorn:2022gku,Biemans:2017zca,Pastor-Gutierrez:2022nki}. Other extensions, incorporating higher-order curvature invariants~\cite{Christiansen:2017bsy,Falls:2017lst,Knorr:2021slg,Kluth:2022vnq,Baldazzi:2023pep}, have reinforced this picture and suggested a finite-dimensional UV critical surface spanned by two or three relevant directions in the gravitational sector~\cite{Falls:2014tra,Falls:2017lst,Kluth:2020bdv}. Investigations on the stability of the fixed point under different gauges and regulator choices~\cite{Nink:2014yya,Gies:2015tca,Knorr:2017fus,DeBrito:2018hur} have strengthened the case for asymptotic safety in QG. Additionally, preliminary investigations seem to indicate that the existence and properties of the fixed point remain unaltered by Wick rotation or within Lorentzian FRG studies~\cite{Manrique:2011jc,Biemans:2016rvp,Biemans:2017zca,Knorr:2018fdu,Eichhorn:2019ybe,Bonanno:2021squ,Fehre:2021eob,Banerjee:2022xvi,DAngelo:2022vsh,DAngelo:2023tis,Pawlowski:2025etp} and that the theory may be unitary~\cite{Bonanno:2021squ,Knorr:2021niv,Fehre:2021eob,Knorr:2024yiu,Eichhorn:2024wba}. The fundamental nature of asymptotic safety may still be challenged by its compatibility with standard black hole thermodynamics~\cite{Basile:2025zjc}, but its impressive consistency with particle physics data~\cite{Shaposhnikov:2009pv, Dona:2013qba,Christiansen:2017cxa,Eichhorn:2017ylw, Eichhorn:2017lry, Eichhorn:2018whv, Eichhorn:2025sux} (see also~\cite{Eichhorn:2022gku} for a review) could point to an effective realization of the asymptotic safety scenario as a predictive bridge between EFT and a more fundamental description~\cite{deAlwis:2019aud}; in this case the fixed point could still be used to explore the landscape of EFTs stemming from a consistent UV completion of gravity~\cite{Held:2020kze,Basile:2021euh,Basile:2021krk,Borissova:2025frj}.

In this context, and despite the achievements of the field, translating the UV properties of the RG flow into macroscopic, semiclassical predictions in the gravitational sector remains a significantly unexplored arena. In the context of black holes, an influential line of work has pursued the idea of ``RG improving'' classical solutions~\cite{Bonanno:1998ye,Bonanno:2000ep,Emoto:2005te,Bonanno:2006eu,Reuter:2006rg,Casadio:2010fw,Falls:2010he,Cai:2010zh,Fayos:2011zza,Falls:2012nd,Koch:2013owa,Bonanno:2016dyv,Bonanno:2017kta,Bonanno:2017zen,Torres:2017ygl,Pawlowski:2018swz,Adeifeoba:2018ydh,Held:2019xde,Platania:2019kyx,Bonanno:2019ilz,Held:2021vwd,Borissova:2022mgd,Eichhorn:2022bgu,Platania:2023srt,Bonanno:2023rzk,Chen:2023pcv,Platania:2025imw}, 
by incorporating the scale dependence of gravitational couplings --- most notably the Newton's and cosmological constants --- into spacetime geometries. In these models, the RG scale is typically identified with an inverse distance measure or a curvature invariant, thereby promoting the couplings to position-dependent quantities. While such improvements have led to important insights, including possible modifications to classical solutions and singularity resolution, the method remains heuristic and sensitive to the choice of scale identification and improvement scheme (however, see~\cite{Borissova:2022mgd} for a possible solution).

The underlying space of quantum effective actions stemming from ASQG is, as expected on general EFT grounds, vastly richer than the Einstein-Hilbert approximation: higher derivative terms are expected to become relevant at high energies and their impact on black hole solutions~\cite{Knorr:2022kqp,Platania:2023uda,Pawlowski:2023dda,Buoninfante:2024oyi} has not been fully charted within a fundamental framework. Nonetheless, recent works~\cite{Lu:2015cqa,Lu:2015psa,Goldstein:2017rxn,Podolsky:2018pfe,Bonanno:2019rsq,Podolsky:2019gro,Silveravalle:2022wij,Bonanno:2022ibv,Silveravalle:2023lnl,Daas:2022iid,Giacchini:2025mlv} have started charting out black holes and their alternatives in the presence of higher derivative operators, in particular in the case of classical quadratic actions as a first step in this direction. For the purpose of this and future work, we introduce the term ``gravitationally localized objects'' (GLOBs) to collectively denote any isolated object --- with or without a horizon, with or without a surface, singular or regular, mimicking or not GR black holes in the limit of large masses --- stemming from the dynamics of a gravity-matter theory. 

In this work, we advance the program of deriving GLOBs from first principles by starting to map out the landscape of static and spherically symmetric spacetimes emerging from an asymptotically safe UV completion. Grounding on existing works~\cite{Bonanno:2019rsq,Silveravalle:2022wij,Silveravalle:2023lnl} we initiate this exploration by focusing on the Einstein-Weyl truncation. In this truncation, the effective action retains not only the Einstein-Hilbert term but also contributions quadratic in curvature, notably the Weyl-squared term. This truncation is particularly instructive, as all asymptotically flat black hole solutions of the general quadratic action are also solutions of the classical Einstein-Weyl theory~\cite{Nelson:2010ig,Lu:2015cqa}. The Wilson coefficients in the effective action are uniquely determined by the UV completion; in this way, all the parameters in the field equations are fixed by the asymptotic safety condition, and the landscape of possible GLOB solutions is obtained by varying the integration constants of the corresponding fourth-order differential equations. We are thus able to chart a GLOB ``phase diagram'' associated with an asymptotically safe UV completion that does not rely on model building and captures a full variety of possible spacetimes. Indeed, as in the general case~\cite{Silveravalle:2022wij}, the phase diagram reveals a rich structure of gravitational solutions, encompassing wormholes, Schwarzschild, and non-Schwarzschild black holes. Notably, asymptotic safety seems to disfavour some types of naked singularities. Our analysis thus provides a first important proof of principle on how to bridge between the microscopic dynamics of ASQG and the landscape of quantum GLOBs, offering a systematic framework to classify and understand quantum gravitational modifications to classical spacetimes. 

The structure of this paper is as follows. In Sect.~\ref{sect:setup}, we motivate the Einstein-Weyl truncation relevant for our analysis. In Sect.~\ref{sect:PDconstr}, we review and revise the phase diagram of Einstein-Weyl gravity derived in~\cite{Silveravalle:2022wij}, refining its consistency domain. Sect.~\ref{sec:ASEW} presents the derivation of the asymptotic safety landscape in the Einstein-Weyl truncation, while Sect.~\ref{sec:BH-from-ASEW} discusses the resulting GLOB phase diagram and the implications of asymptotic safety for the latter. Finally, in Sect.~\ref{sec:conclusions}, we summarize our findings and outline directions for future research.

\section{Preliminaries}\label{sect:setup}

In this section, we introduce the general idea behind our work, as well as its concrete setup.

\subsection{Charting black holes and their alternatives from quantum gravity}\label{sec:ChartingBH}

\begin{figure}[t!]
	\centering
	\includegraphics[width=0.65\textwidth]{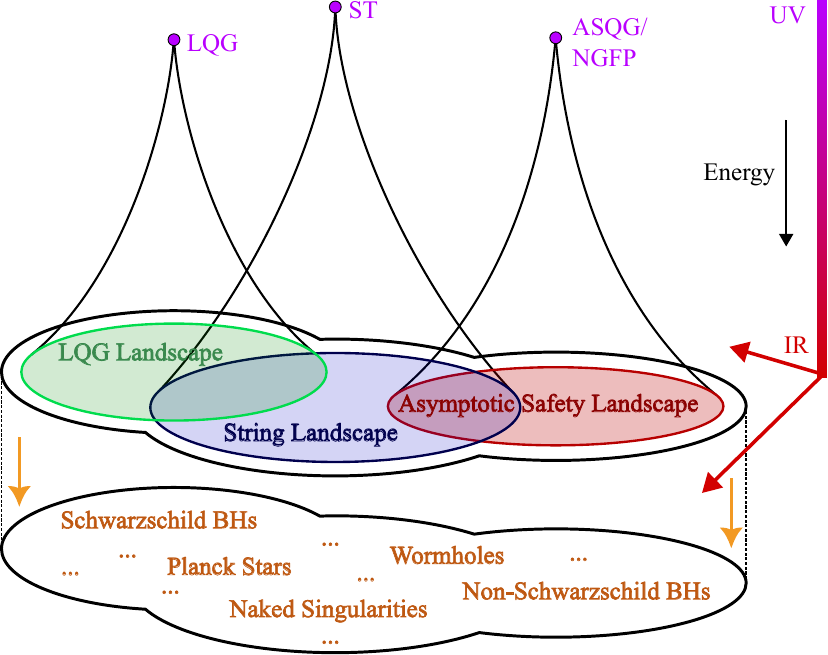}
	\caption{Sketch of the idea behind our work. The set of viable EFTs stems from different UV completions. The resulting quantum effective actions can then be mapped onto different GLOBs by solving the corresponding field equations.}
\label{fig:landscape}
\end{figure}

Different approaches to QG are expected to correspond to distinct UV completions of the gravitational dynamics. In the light of the spectrum of possible UV completions~\cite{Bambi2024-nm,Buoninfante:2024yth}, one may adopt a theory-agnostic, operational approach by characterizing the low-energy imprints of different UV completions in terms of sets of effective actions: each UV completion is mapped onto a landscape of EFTs (cf. Fig.~\ref{fig:landscape}).
The EFT is organized as a derivative expansion of diffeomorphism-invariant operators,
\begin{equation}
\Gamma = \int d^4x \, \sqrt{-g} \left( \frac{m_{\mathrm{Pl}}^2}{2} R + \sum_i c_i \, \mathcal{O}_i[g, \phi] \right),
\end{equation}
where the operators $\mathcal{O}_i$ include higher-derivative terms, couplings to additional fields, collectively called $\phi$, or nonlocal structures, depending on the UV scenario. The Wilson coefficients $\{c_i\}$ vary across the landscape induced by the underlying QG theory, and can be used to parametrize it --- at least in the regime where QFT holds.

The approach we pursue here is to treat this landscape as the arena in which to derive and classify gravitational solutions, in particular static and spherically symmetric ones. Each set of coefficients $\{c_i\}$ corresponds to an effective action and hence to specific field equations. The corresponding solutions may or may not admit an event horizon, may be singular or regular, and can include small or large deviations from classical Schwarzschild-like behavior at both short and long distances. The diversity of UV completions is thus expected to translate into a variety of generalized compact objects, each tied to a specific region of the EFT parameter space. In this framework, GR corresponds to the origin, $c_i=0$, corresponding to Schwarzschild black holes. For $c_i\neq0$, and in particular once these Wilson coefficients are fixed by a specific UV completion, the space of solutions can be explored by solving the resulting field equations. Rather than yielding a single alternative to classical black holes, generic QG corrections are expected to give rise to a landscape of GLOBs, which may include wormholes, regular black holes, or other types of isolated objects. Their detailed features depend both on the EFT parameters (i.e., the specific point in the QG landscape) and on macroscopic integration constants such as the mass $M$, spin $J$, and possibly other scalar charges. Hence, once the Wilson coefficients are fixed, the different solutions can be mapped out in a ``phase diagram'' whose axes are integration constants, like the object's mass. This framework thus provides a bridge between QG and the possible GLOBs emerging from specific UV scenarios. We remark that for theories where QFT does not hold at all scales, the approach would still allow to map out GLOBs whose size is much larger than the Planck volume, but may not give access to all spacetime solutions of the theory.

In this work, we provide a proof of concept of the idea explained in this paragraph, focusing on asymptotic safety as a UV completion, and Einstein-Weyl gravity as an approximation to the dynamics. In the following, we provide more details on the latter.

\subsection{Setup: Einstein-Weyl truncation}

Our investigations focus on the Einstein-Weyl truncation of the effective action,
\be \label{eq:EWaction}
\Gamma_{\rm EW}= \frac{1}{16 \pi \GN} \int d^4x \,\sqrt{-g} \left( R - \frac{1}{2} \GC \, C^2 \right)\,,
\ee
where $C^2= C_{\mu \nu \rho \sigma} C^{\mu \nu \rho \sigma}$ and $C_{\mu \nu \rho \sigma}$ is the Weyl tensor. This is clearly a very simplified approximation of the full effective action, retaining only two key terms: the Einstein-Hilbert term, which is the leading-order term in the gravitational EFT, and the Weyl-squared term~$C^2$. This term leads to a ghost degree of freedom and hence to a violation of unitarity~\cite{Stelle:1977ry}. However, our perspective is that~\eqref{eq:EWaction} is the leading-order correction in a standard EFT expansion, and that once a momentum dependence of operators has been reconstructed, the ghost pole would be removed. Preliminary studies suggest that this is the case in ASQG~\cite{Bonanno:2021squ,Fehre:2021eob,Knorr:2021niv}, so that the ghost pole in approximations like the one above would simply be a truncation artifact which decouples when the truncation order is increased~\cite{Platania:2020knd,Platania:2022gtt}. The motivation for this truncation lies in balancing tractability with a richer dynamics: the Weyl-squared term encodes some of the leading-order higher-derivative corrections that are part of the EFT, the truncation is relatively simple to analyze via RG techniques and yield an approximation to the fixed point of ASQG~\cite{Knorr:2021slg}, and the system offers a richer spectrum of solutions, that is the only one whose phase diagram has been throughly studied~\cite{Silveravalle:2022wij}. The Einstein-Weyl truncation thus serves as a minimal yet nontrivial framework to explore quantum gravitational effects in black hole physics. 

\section{Phase diagram of classical Einstein-Weyl gravity}\label{sect:PDconstr}

We consider the space of static, spherically symmetric vacuum solutions of Einstein-Weyl theory, described by the metric ansatz
\begin{equation}\label{eq:metric}
    \mathrm{d}s^2=-h(r)\mathrm{d}t^2+\frac{\mathrm{d}r^2}{f(r)}+r^2\mathrm{d}\Omega^2\,.
\end{equation}
By imposing asymptotic flatness at spatial infinity, one can analytically derive a weak-field expression for the metric from the linearized equations of motion (EOM) at large distances~\cite{Stelle:1977ry,Bonanno:2019rsq}
\begin{equation}\label{eq:wfmetric}
    \begin{split}
        & h(r)\sim 1-\frac{2 \GN M}{r}+2 S_2^-\frac{\mathrm{e}^{-m_2 r}}{r},\\
        & f(r)\sim 1-\frac{2 \GN M}{r}+S_2^-\frac{\mathrm{e}^{-m_2 r}}{r}\left(1+m_2 r\right)\,,
    \end{split}
\end{equation}
where $m_2=1/\sqrt{\GC}$, $M$ is the total ADM mass\footnote{Note that, while they generally converge asymptotically, in higher derivative theories one could define different notions of masses~\cite{AparicioResco:2016xcm,Astashenok:2017dpo,Sbisa:2019mae,Bonanno:2021zoy}.} and $S_2^-$ is a free parameter which we will call Yukawa charge from now on. Once the parameters of the theory $\GN$ and $m_2$ are fixed, the mass $M$ and the Yukawa charge $S_2^-$ completely characterize the external gravitational field of the solution and can therefore be regarded as its \qu{gravitational state parameters}. A chart representing different classes of solutions with distinct qualitative features (such as the presence of an event horizon or a wormhole throat) in terms of these gravitational state parameters can be referred to as the \qu{phase diagram} of the theory. This is by analogy with standard phase diagrams, where different phases of matter are plotted in terms of macroscopic parameters: the thermodynamic state variables. The analogy is, however, pictorial: it is not clear, for instance, if one could start from one solution and transition to another one by varying the gravitational state parameters.

\subsection{Current understanding of the phase diagram}

Extensive analytical and numerical analysis clarified that there are four different types of static, spherically symmetric vacuum solutions (in addition to the trivial Minkowski vacuum) \cite{Lu:2015cqa,Lu:2015psa,Goldstein:2017rxn,Podolsky:2018pfe,Bonanno:2019rsq,Podolsky:2019gro,Silveravalle:2022wij,Bonanno:2022ibv,Silveravalle:2023lnl,Daas:2022iid}, of which we briefly sketch their main properties.

\paragraph{Black holes~\cite{Lu:2015cqa,Goldstein:2017rxn,Bonanno:2019rsq}.} They are characterized by the presence of an event horizon, which is a simultaneous root of the metric functions $h(r)$ and $f(r)$, and can be present with both Schwarzschild and non-Schwarzschild metrics. In the case of the latter, in contrast to standard solutions, as the radius of the event horizon decreases, their total mass and entropy increase to finite values, while their temperature approaches zero. Additionally, non-Schwarzschild black holes with a horizon radius smaller than a specific value $\bar{r}_H$ have a positive Yukawa charge and a vanishing metric at the origin, while those with a horizon radius larger than $\bar{r}_H$ have a negative Yukawa charge and a divergent metric at the origin. For linear perturbations, both Schwarzschild and non-Schwarzschild black holes are unstable when the radius is smaller than $\bar{r}_H$ \cite{Lu:2017kzi,Held:2022abx,Silveravalle:2023lnl,East:2023nsk,Antoniou:2024jku}.

\paragraph{Type I exotic solutions: repulsive naked singularities~\cite{Lu:2015psa,Silveravalle:2022wij,Silveravalle:2023lnl}.} They are characterized by a divergent behavior of the metric near the origin, corresponding to a repulsive gravitational potential, similar to standard naked singularities (e.g., Reissner-Nordstr\"om). This type of singularity cannot be reached by a massive particle and provides an extreme blueshift to photons emitted close to it. They are not valid black hole mimickers, as they do not have a shadow and an orbiting gas emits photons with extremely high energies~\cite{Daas:2022iid}. A preliminary analysis of their behavior under dynamical perturbations suggests that they are stable only when the mass is sufficiently larger than the Yukawa charge~\cite{Silveravalle:2023lnl}.

\paragraph{Type II exotic solutions: attractive (Bachian) naked singularities~\cite{Lu:2015psa,Silveravalle:2022wij,Silveravalle:2023lnl,Holdom:2016nek,Holdom:2022zzo}.} They are characterized by a vanishing metric near the origin, which corresponds to an attractive gravitational potential. They are a feature of quadratic gravity and lack similar counterparts in GR. A free-falling object will reach the singularity in a finite time, and as it gets closer, it will experience extreme radial tidal forces. In contrast to repulsive naked singularities, photons moving away from the singularity experience an extreme redshift and reach spatial infinity with very low frequencies. They are valid black hole mimickers, as they have a shadow slightly smaller than that of a Schwarzschild black hole with the same mass. Due to their attractive nature, they appear to be stable under linear perturbations from a preliminary analysis~\cite{Silveravalle:2023lnl}.

\paragraph{Type III exotic solutions: non-symmetric wormholes~\cite{Lu:2015psa,Silveravalle:2022wij,Silveravalle:2023lnl,Bonanno:2022ibv}.} They are characterized by the presence of a wormhole throat, which is a root of the function $f(r)$ but not of $h(r)$. They connect an asymptotically flat spacetime to a singular universe of finite size. The gravitational force is attractive towards the throat in the asymptotically flat patch and repulsive in the other patch. Similar to attractive naked singularities, free-falling objects will experience extreme tidal forces close to the singularity, and photons that move towards spatial infinity in the asymptotically flat patch will be highly redshifted. They are valid black hole mimickers, as they have a shadow slightly larger than that of a Schwarzschild black hole with the same mass. The preliminary analysis of linear perturbations suggests that they are always unstable, which is a common behavior for traversable wormhole solutions~\cite{Silveravalle:2023lnl}.

\begin{figure}[t]
    \centering
\includegraphics[width=0.7\textwidth]{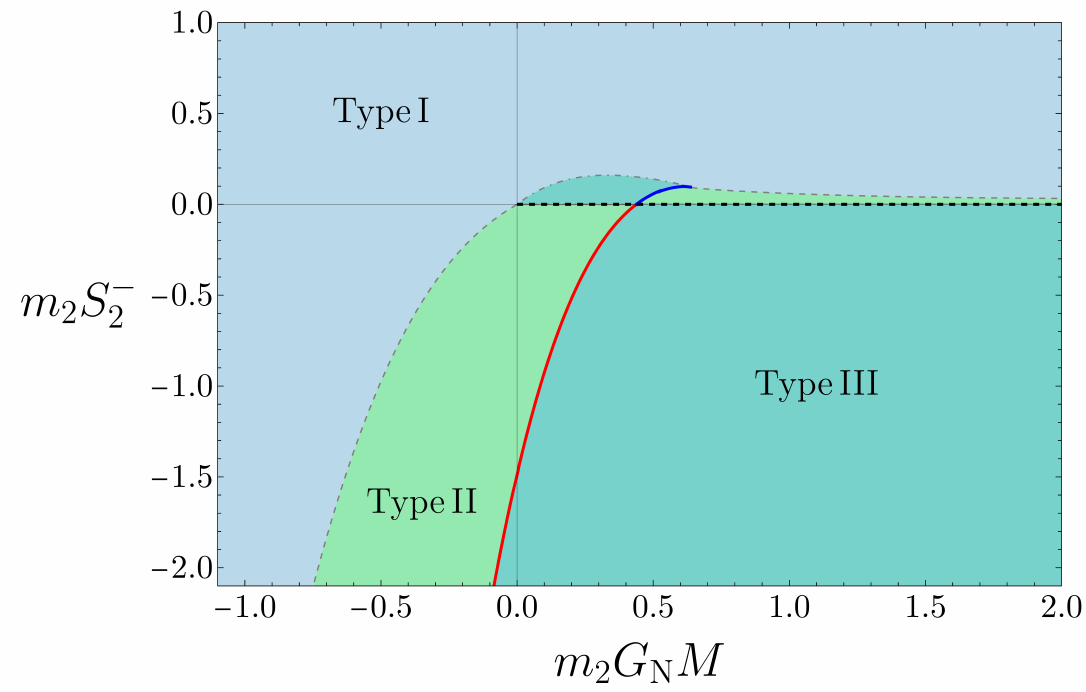}
    \caption{The phase diagram of static, spherically symmetric solutions in Einstein-Weyl gravity, as presented in~\cite{Silveravalle:2022wij}, but with the dependence on $m_2$ explicitly shown in the axes. The axes show the mass $M$ and Yukawa charge $S_2^-$, both made dimensionless using combinations of $m_2$ and $\GN$. The dashed black line represents Schwarzschild black holes, while the solid red and blue lines represent non-Schwarzschild ones. The dotted and dash-dotted gray lines indicate approximate parameter values where smooth transitions between Type I and other solution types occur.}
    \label{fig:pd-review}
\end{figure}

In Fig.~\ref{fig:pd-review}, we show the phase diagram of Einstein-Weyl gravity, as presented in \cite{Silveravalle:2022wij}. Black holes appear in zero-measure regions of the parameter space, that is, on the lines separating attractive naked singularities from non-symmetric wormholes. The two lines corresponding to Schwarzschild and non-Schwarzschild black holes cross each other at a specific point, given by $m_2 \GN \bar{M} = m_2\bar{r}_H/2 \simeq 0.438$ and $m_2 \bar{S}_2^- = 0$, and end respectively in the Minkowski vacuum and a ``massive triple point'' located at $m_2 \GN M_{\text{mtp}} \simeq 0.626$ and $m_2 S_{2,\text{mtp}}^- \simeq 0.097$. Attractive naked singularities and wormholes both transition smoothly into repulsive naked singularities when the mass becomes sufficiently large and negative, or when the Yukawa charge becomes sufficiently large and positive. In analogy with standard phase diagrams, we may interpret black holes as phase transitions between attractive naked singularities and wormholes, the Minkowski vacuum and the massive triple point as critical points, and repulsive naked singularities as a supercritical state.

In the phenomenologically interesting regime of large masses, which in our case means $M \gg m_{\rm Pl}^2/m_2$, black holes can only be those characterized by the Schwarzschild metric. A negative Yukawa charge leads to a non-symmetric wormhole solution, while small positive Yukawa charges ($0< S_2^- \ll 1/m_2$) yield attractive naked singularities. For larger positive values, the solutions correspond to repulsive naked singularities. The only other non-trivial regime is that of large, negative Yukawa charges. In this case, black holes can only have non-Schwarzschild metrics and negative masses. All solutions with positive mass correspond to non-symmetric wormholes, while for negative masses, the solutions correspond to wormholes, attractive naked singularities, and repulsive naked singularities, respectively, as the absolute value of the mass increases.

While it is interesting that such a rich phase diagram arises from the simple addition of a single term to the Einstein-Hilbert action, its interpretation may benefit from constraints on the allowed range of parameters, justified by physical or mathematical arguments. Novel constraints of this type are addressed in the next subsection.

\subsection{Revisiting the weak field expansion and self-consistency}\label{sec:RegionofValidity}

The weak-field expression for the metric in Eq.~\eqref{eq:wfmetric} has a clear physical meaning as the sum of the gravitational potentials of a massless and a massive particle. Nonetheless, its derivation from a linearized expansion requires some care when assessing its domain of validity. A strong consistency condition for the linearization procedure is that all the terms in Eq.~\eqref{eq:wfmetric} should be of the same order and much smaller than unity: 
\begin{equation}\label{eq:consistency}
    \frac{\GN M}{r}\sim S_2^-\frac{\mathrm{e}^{-m_2 r}}{r}\sim S_2^- m_2 \mathrm{e}^{-m_2 r}\ll 1\,.
\end{equation}
The first consequence is that Eq.~\eqref{eq:wfmetric} is valid in the region of spacetime where $r\sim m_2^{-1}$, which is unsurprising, given that for $r\gg m_2^{-1}$ the exponential terms are effectively zero. The second consequence is that the gravitational state parameters of the solutions must be of the same order and much smaller than the relevant length scale: $\GN M\sim S_2^-\ll m_2^{-1}$. However, we know that the Schwarzschild metric is a solution of the full non-linear EOM, so no constraint should be imposed on $M$.

To perform a consistency check that takes into account the special role of the Schwarz\-schild solution, we begin by considering a linear expansion around a Schwarzschild background:
\begin{equation}\label{eq:wfschwa}
    \begin{split}
        & h(r)\sim 1-\frac{2 \GN \longbar{M}}{r}+\epsilon\,V(r),\\
        & f(r)\sim 1-\frac{2 \GN \longbar{M}}{r}+\epsilon\,W(r)\,.
    \end{split}
\end{equation}
A solution to the EOM with the metric ansatz in Eq.~\eqref{eq:wfschwa} will be a small deviation from the global Schwarzschild metric, and must thus ensure the presence of an event horizon. Such a solution is known~\cite{Lu:2017kzi} and can only be realized at the point in parameter space where the two branches of Schwarzschild and non-Schwarzschild black holes intersect. Nonetheless, a possible refinement of the weak-field expansion can be sought by considering a double expansion:
\begin{equation}\label{eq:wfdouble}
    \begin{split}
        & h(r)\sim 1-\frac{2 \GN \longbar{M}}{r}+\epsilon\left(V^{(0)}(r)+V^{(1)}(r)+V^{(2)}(r)+V^{(3)}(r)+V^{(4)}(r)\right),\\
        & f(r)\sim 1-\frac{2 \GN \longbar{M}}{r}+\epsilon\left(W^{(0)}(r)+W^{(1)}(r)+W^{(2)}(r)+W^{(3)}(r)+W^{(4)}(r)\right),
    \end{split}
\end{equation}
where the superscripts ${}^{(n)}$ indicate that the functions satisfy the EOM expanded up to order $O\left(\left(\frac{\GN \longbar{M}}{r}\right)^n\right)$. While the zeroth-order EOM can be fully solved analytically in Fourier space, with Eq.~\eqref{eq:wfmetric} as the result, all higher-order terms are not exactly solvable. Nonetheless, knowing the form of $V^{(0)}(r)$ and $W^{(0)}(r)$ we can look for solutions using a Frobenius-like method \cite{Lu:2015psa,Podolsky:2018pfe} with the ansatz
\begin{equation}\label{eq:wffrob}
    \begin{split}
        & V^{(n)}(r)= \mathrm{e}^{-c\, m_2 r}r^a\displaystyle\sum_{i=0}^{N}\frac{h_{i+a}}{r^i}+O\left(\frac{1}{r^{N+1}}\right),\\
        & W^{(n)}(r)= \mathrm{e}^{-c\, m_2 r}r^b\displaystyle\sum_{i=0}^{N}\frac{f_{i+b}}{r^i}+O\left(\frac{1}{r^{N+1}}\right).
    \end{split}    
\end{equation}
The solution, perhaps unsurprisingly, is
\begin{equation}\label{eq:wfmetricrev}
    \begin{split}
        & h(r)\sim 1-\frac{2 \GN \left(\longbar{M}+M^{(0)}+M^{(1)}+M^{(2)}+M^{(3)}+M^{(4)}\right)}{r}+2 S_2^{-(0)}\frac{\mathrm{e}^{-m_2 r}}{r},\\
        & f(r)\sim 1-\frac{2 \GN \left(\longbar{M}+M^{(0)}+M^{(1)}+M^{(2)}+M^{(3)}+M^{(4)}\right)}{r}+S_2^{-(0)}\frac{\mathrm{e}^{-m_2 r}}{r}\left(1+m_2 r\right),
    \end{split}
\end{equation}
which reduces to Eq.~\eqref{eq:wfmetric} after defining $M=\longbar{M}+M^{(0)}+M^{(1)}+M^{(2)}+M^{(3)}+M^{(4)}$ and $S_2^-=S_2^{-(0)}$. What we would like to stress, however, is that here the mass parameter $\longbar{M}$ does not come from a linear expansion and is therefore not subject to any constraint. The consistency requirement in Eq.~\eqref{eq:consistency} thus becomes
\begin{equation}\label{eq:consistencyrev}
    \frac{\GN M^{(n)}}{r}\sim S_2^{-(n)}\frac{\mathrm{e}^{-m_2 r}}{r}\sim S_2^{-(n)} m_2 \mathrm{e}^{-m_2 r}\ll \left(\frac{\GN \longbar{M}}{r}\right)^n.
\end{equation}
The two consequences are, once again, that the asymptotic expansion in Eq.~\eqref{eq:wfmetricrev} is valid where $r\sim m_2^{-1}$, and that $\GN (M-\longbar{M})\sim S_2^-\ll m_2^{-1}$; in other words, the phase diagram in Fig.~\ref{fig:pd-review} is reliable only when the gravitational state parameters $M$ and $S_2^-$ do not deviate significantly from those of a Schwarzschild black hole. Given that this consistency constraint is quite rigid in its formulation, we do not impose a sharp cutoff in the parameter space, but rather consider a smooth increase in the level of uncertainty beyond a distance of $m_2^{-1}$ from the Schwarzschild region of parameters. Therefore, understanding the value of $m_2$, and thus setting the scales of the solutions, is crucial for the physical interpretation of the phase diagram. Seeing Einstein-Weyl gravity as a truncation of the standard gravitational EFT expansion, $m_2$ is a Wilson coefficient whose value depends on the specific UV completion of gravity. In the next section, we shall focus on one particular UV completion, namely, ASQG, where the calculation of $m_2$ is feasible.

\section{UV-completing Einstein-Weyl gravity with asymptotic safety} \label{sec:ASEW}

The phase diagram we discussed in the previous section uses the Einstein-Weyl action as a simple beyond-GR model, with free unconstrained couplings. Eventually, some of these couplings (or Wilson coefficients) ought to be predicted by a fundamental QG theory. The idea of this work is to embed the phase diagram in a fundamental QFT framework and constrain some parts of it. This requires two ingredients:
\begin{itemize}
    \item To treat the Einstein-Weyl action as a truncation of the full EFT expansion, so that the additional degree of freedom can be fictitious in the sense of~\cite{Platania:2020knd} and not necessarily a ghost.
    \item To find a UV completion for the Einstein-Weyl action, which is an approximation to the Reuter fixed point of ASQG.
\end{itemize}
One can then use this UV completion to compute the asymptotic safety landscape~\cite{Basile:2021krr,Knorr:2024yiu}, predict the Weyl coupling $\GC$, and constrain the GLOB phase diagram of~\cite{Silveravalle:2022wij}. 

\subsection{From the UV to the asymptotic safety landscape} \label{s:RGflowanalysis}

The RG flow of Einstein-Weyl gravity is a special case of quadratic gravity, where operators up to quartic order in derivatives are considered. The beta functions of this system have been derived in~\cite{Knorr:2021slg} via the FRG. We shall use them for our analysis, which consists of (i) finding the fixed point for this system, and (ii) deriving the asymptotic safety landscape stemming from it, thereby setting a bound on the Wilson coefficient $\GC$. To specialize to the case of the Einstein-Weyl truncation, we need to project the flow of~\cite{Knorr:2021slg} on the subspace in which only the couplings $\GC$ and $\GN$ are kept generically nonzero.
Therefore, the beta functions of~\cite{Knorr:2021slg} can be adapted to our case, defining a bi-dimensional flow determined by two coupled differential equations
\be \label{eq:def_beta}
\beta_g[g,\gc] := \frac{\rmd g}{\rmd \tau} \,, \qquad \beta_{\gc}[g,\gc] := \frac{\rmd \gc}{\rmd \tau} \,.
\ee
Here we have introduced the \qu{RG time} $\tau=\log (k/k_0)$, where $k_0$ is a reference scale which emerges as an integration constant and has the interpretation of a transition scale to the fixed-point regime. Moreover, as it is standard in FRG computations, we have introduced dimensionless versions of the interaction couplings by rescaling them with appropriate powers of the RG scale $k$,
\be \label{eq:dimensionless_couplings}
g(k) := \GN(k) k^2 \,, \qquad \gc(k) := \GC(k) k^2 \,.
\ee

Let us stress here that, in terms of a path integral formulation, $k$ plays the role of an infrared (IR) cutoff for the integration of quantum fluctuations. Therefore, the limit $k\to 0$ (or, equivalently, $\tau \to - \infty$) corresponds to performing a functional integral over all the quantum fluctuations and is tantamount to computing the quantum effective action of a theory \cite{Percacci:2017fkn}. 
In this subsection, we detail the procedure we followed to compute the asymptotic safety landscape~\cite{Basile:2021krr,Knorr:2024yiu}, whereas in the next subsection, we will present the quantitative results of this analysis.

The integration of Eq.~\eqref{eq:def_beta} yields the flow of the theory from the  UV ($k \to \infty$) to the IR ($k \to 0$). If the theory is asymptotically safe, it shows scale invariance in the UV: as~$k$ grows, the couplings approach a NGFP $(g_\ast, \,g_{\rm C^2,*}) \ne (0,\, 0)$ at which, by definition, the beta functions vanish,
\be \label{eq:FPdef}
\beta_g[g_*,\, g_{\rm C^2,*}] =0 \,, \qquad \beta_{\gc} [g_*,\, g_{\rm C^2,*}] =0 \,.
\ee
If such a UV completion exists, one can ask which EFTs stem from it. In particular, the predictions on the values of the Wilson coefficients can be obtained by considering all the possible IR endpoints of the trajectories $(g(k), \,  \gc(k))$ which depart from the NGFP in the UV. This defines the asymptotic safety landscape~\cite{Basile:2021krr,Knorr:2024yiu}.

The extraction of the asymptotic safety landscape and the computation of the Wilson coefficients can be done following the steps in~\cite{Basile:2021krr,Knorr:2024yiu}. In the IR limit, the theory approaches the GR limit, where the flow is controlled by the Gaussian fixed point (GFP) and the couplings follow a canonical scaling. Moreover, the presence of massless fluctuations (the graviton in our case) can generate a logarithmic IR running in $\gc(k)$~\cite{Basile:2021krr,Knorr:2024yiu}. Therefore, in the IR limit, we expect a universal behavior of the form
\be \label{eq:IR_scaling}
g(k)=g_{\rm IR} \left( \frac{k}{k_0} \right)^2 \,, \qquad \gc(k) = \gci \left( \frac{k}{k_0} \right)^2 \left[ 1 + b \log\left( \frac{k}{k_0} \right) \right] \,,
\ee
where $\{g_{\rm IR}, \, \gci, \, b \}$ are constants. One can immediately notice that the logarithmic term generates an ambiguity in the determination of the IR limit of $\gc(k)$: a change in the (arbitrary) scale $k_0$ reflects into a change of the value of $\gci$, while leaving the product $b\gci$ unaffected. Defining Wilson coefficients in the presence of massless fluctuations is a general problem in the EFT literature, and in the context of ASQG it has been extensively discussed in~\cite{Basile:2021krr,Knorr:2024yiu}, where a prescription to fix the scale and subtract the logarithmic running and define the Wilson coefficient has been introduced. In the following, we detail this procedure. 

Let us first focus on the individual runnings. From Eq.~\eqref{eq:dimensionless_couplings}, one can, by definition, fix Newton's coupling as the IR limit of $\GN(k)$
\be
\GN = \lim_{k \to 0} \GN (k)=\lim_{k \to 0} g(k) k^{-2}= g_{\rm IR} k_0^{-2} \,.
\ee
Hence, in the IR limit, one can also rewrite $g(k)=\GN k^2$.
At this point, one can use the definition of $\GN\equiv m_\text{Pl}^{-2}$ to set both the units and the transition scale $k_0$. This is equivalent to setting $g_{\rm IR}=1$ and $k_0^2= m_\text{Pl}^2$. With this choice, in the IR $\gc(k)$ reads 
\be
\gc(k)=\gci \GN k^2 \left[ 1  +   b \log\left( k/m_\text{Pl} \right) \right]\,.
\ee
The last two equations define the running of the dimensionless Newton's and Weyl-squared couplings. In general, for any given gravity-matter system, one dimensionful coupling has to be used to set the unit scale of the system, and hence to define dimensionless ratios: these are the only meaningful physical quantities that one can construct. In our case, we choose Newton's coupling, i.e., Planck units, to set the unit scale of the system and to define dimensionless Wilson coefficients. The only dimensionless Wilson coefficient in our system is $\GC/\GN$, and this is also the relevant one to construct the GLOB phase diagram in ASQG. A crucial observation is that
\be
\frac{\GC}{\GN}=\lim_{k\to0}\frac{\GC(k)}{\GN(k)}\equiv\lim_{k\to0}\frac{\gc(k)}{g(k)} \,,
\ee
since $\GC$ and $\GN$ have the same classical mass dimension. 
To compute the relevant dimensionless Wilson coefficient, we thus need to consider the ratio
\be \label{eq:ratio_IR}
\frac{\gc(k)}{g(k)}=\gci \left[ 1  +   b \log\left( k /m_\text{Pl} \right) \right]= \gci  \left( 1 + b \, \tau \right) \,,
\ee
where now we have chosen~$k_0= m_\text{Pl}$ so that, by definition,  the RG time vanishes at the Planck mass, $\tau(m_\text{Pl})=0$. Once the scale $k_0$ is fixed, we have a well-defined prescription to single out the value of $\gci$. Following~\cite{Basile:2021krr,Knorr:2024yiu}, this is done by subtracting the logarithm in Eq.~\eqref{eq:ratio_IR},
\be \label{eq:prescription_G_N}
\gci= \lim_{k \to 0}\left[ \frac{\gc(k)}{g(k)} - \left( k \frac{\partial}{\partial k } \frac{\gc(k)}{g(k)} \right) \log\left(  k/m_\text{Pl} \right)\right] \,.
\ee
Here the slope $ k \frac{\partial}{\partial k } \frac{\gc(k)}{g(k)}$ is equivalent to $b$ in Eq.~\eqref{eq:ratio_IR}.
With this subtraction, the dimensionful Weyl-squared Wilson coefficient is given by
\be \label{eq:alphaIR}
\GC=\gci \GN \,.
\ee
We remark that the prescription~\eqref{eq:prescription_G_N} depends on the scale $k_0$ and on $g_\text{IR}$: while, by definition, $\GN\equiv g_\text{IR} k_0^{-2}$, the scale $k_0$ is arbitrary, and using a different scale $\tilde{k}_0 \ne m_\text{Pl}$ would shift the value of $\gci$ to~\cite{DelPorro:2025fiu}
\be 
\gci|_{\tilde{k}_0}=\gci|_{k_0=m_{\text{Pl}}}\left[1 - b \log \small(\tilde{k}_0/m_\text{Pl}\small) \right] \,.
\ee
This would be equivalent to extracting a Wilson coefficient in units of $\tilde{k}_0$, $\GC=\gci(\tilde{k}_0) \tilde{k}_0^{-2}$. We will comment later on how this would affect our results. At the same time, while $k_0$ is arbitrary, if it were very different from the Planck mass, it would signify that the system develops an additional transition scale, unrelated to the Planck mass. This may be possible in a scenario of effective ASQG~\cite{deAlwis:2019aud,Basile:2021euh,Basile:2021krk}, where the flow stems from a more fundamental theory, characterized by an additional scale like the string mass; yet, in a scenario of fundamental ASQG, as the one we are considering here, the Planck scale is the only relevant scale and hence it is natural to use it to subtract the logarithm and define the Wilson coefficient. With this prescription, every quantity is measured in Planck units and the Planck mass also sets the scale of QG. 

\subsection{RG flow analysis and prediction for the Weyl coupling from ASQG} \label{sec:prediction}

\begin{figure}[t]
    \centering
\includegraphics[width=0.5\textwidth]{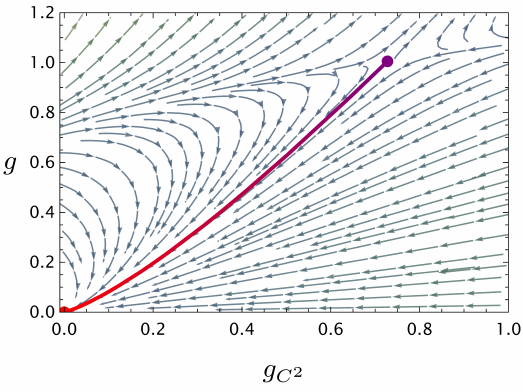}
    \caption{RG flow of the theory in the $\{g, \, g_{\text{C}^2}\}$ plane, parametrized by the RG time~$\tau=\log k/k_0$ and based on an appropriated projection of the beta functions of~\cite{Knorr:2021slg}. The arrows point in the IR direction. There is a unique UV-complete trajectory (solid purple-to-red line), which is also the separatrix connecting the NGFP to GFP.}
    \label{fig:traj}
\end{figure}

Armed with the theoretical background of the section above, we are now ready to analyze the RG flow of the Einstein-Weyl action in Eq.~\eqref{eq:EWaction} and make a prediction for the Weyl-squared Wilson coefficient. The projection of the RG flow of~\cite{Knorr:2021slg} is done by setting the cosmological constant and Ricci-squared couplings to zero. The resulting beta functions are numerical, and the fixed point condition in Eq.~\eqref{eq:FPdef} gives us --- as expected in the context of ASQG --- two solutions
\be
(g_*, \, g_{\text{C}^2,  *})_{\rm GFP}=\left(0, \, 0 \right) \,, \quad \mbox{and} \quad (g_*, \, g_{\text{C}^2,  *})_{\rm NGFP}=(1.0053, \, 0.7277) \,.
\ee
The former is the GFP, while the latter is an NGFP. To characterize the NGFP, we analyze the stability matrix $\mathcal{M}_{ij}\equiv\partial \beta_i/\partial g_j$, with $g_i=\{g,g_{\text{C}^2}\}$.
The eigenvalues of $\mathcal M$ at the NGFP give us the two critical exponents
\be
\theta_1=2.6165 \,, \quad \mbox{and} \quad \theta_2=-0.9365 \,.
\ee
The opposite signs of the two critical exponents tell us that the NGFP is a saddle point: it is UV-attractive along one direction in the $\{g, \, g_{\text{C}^2}\}$ plane and UV-repulsive in the other, respectively given by the two eigenvectors of $\mathcal M|_{\rm NGFP}$:
\be
v_1=(-0.8654,-0.5011) \,, \quad \mbox{and} \quad v_2=(-0.2569,-0.9664) \,.
\ee
In general, relevant directions correspond to free parameters, while the irrelevant ones yield predictions for the Wilson coefficients in the effective action. One of the relevant couplings ought to set the units, and hence, a theory stemming from a QFT-based UV completion with $N$ relevant directions will lead to a landscape of EFTs parametrized by $N-1$ free parameters.
In particular, in our case, this implies that the theory has zero free parameters: the asymptotic safety landscape resulting from the NGFP is a single EFT whose Weyl-squared Wilson coefficient is predicted in Planck units by the unique UV-complete trajectory departing from the NGFP along $v_1$. In the IR, this trajectory will hit the GFP, which is completely IR attractive. In Fig.~\ref{fig:traj} we show the flow of the theory in the bi-dimensional space spanned by the two couplings $\{g, \, g_{\text{C}^2}\}$, highlighting the UV-complete trajectory and the fixed points.

\begin{figure}[t!]
	\centering
	\includegraphics[width=0.55\textwidth]{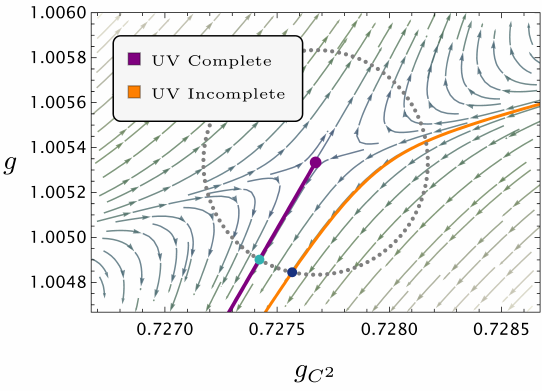}
	\caption{The figure shows a zoom of the RG flow of the theory in the proximity of the NGFP and illustrates the method for identifying the UV-complete RG trajectory connecting the two fixed points (purple line). The initial condition for this separatrix ought to lie on the relevant eigendirection of the NGFP, which points towards the GFP. However, the linearization identifying the eigendirection is only valid in the proximity of the NGFP, while away from it, this relevant line bends to meet the GFP in the IR. Hence, if the initial point is not sufficiently close to the NGFP, despite it lying on the linearized relevant direction, the corresponding RG trajectory may not correspond to the UV-complete one. In other words, starting from a slightly wrong initial condition (e.g., the dark blue dot), the RG trajectory solving the beta functions (orange line) would not hit the NGFP and shoot off to infinity. Tuning the initial condition to be very close to the NGFP and on the right side of the relevant eigendirection (light blue dot), one can cross-check that integrating the flow ``upwards'' (towards the UV), the corresponding RG trajectory hits the NGFP in the UV --- up to numerical precision. Once the UV-complete trajectory is found, one can use the same initial condition to integrate the flow downwards and extract the IR limit of the flow. This then allows us to determine the Wilson coefficients of the theory.}
\label{fig:findNGFP}
\end{figure}

We are now ready to integrate the projected RG flow to extract the prediction for the Weyl-squared Wilson coefficients. To this end, a first crucial step is identifying the right initial condition corresponding to the asymptotically safe RG trajectory. Indeed, while it may seem sufficient to move one step away from the NGFP along a relevant eigendirection, if this step is not ``small enough'', the linearization of the flow about the NGFP identifying its characteristic eigendirections breaks down. Hence, it is crucial to take the numerical initial condition sufficiently close to the NGFP and to cross-check, by integrating the flow upward, that the trajectory indeed hits the NGFP within numerical precision. This process is illustrated in Fig.~\ref{fig:findNGFP}.

\begin{figure}[t!]
    \centering
    \begin{subfigure}[b]{0.51\linewidth}
        \centering
        \includegraphics[width=\linewidth]{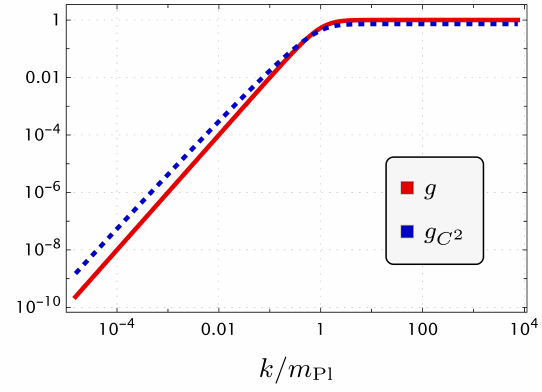}
    \end{subfigure}
    \hfill
    \begin{subfigure}[b]{0.48\linewidth}
        \centering
        \includegraphics[width=\linewidth]{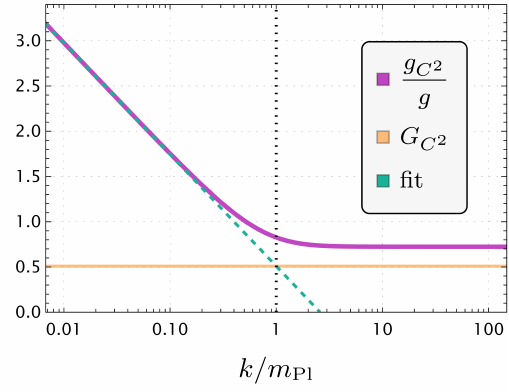}
    \end{subfigure}
    \caption{Running of the couplings $\{g, \gc\}$ (left panel) and their ratio (right panel) in the Einstein-Weyl truncation, as a function of the RG scale $k$, in Planck units. The left panel illustrates how $g(k)$ (red solid line) and $\gc(k)$ (blue dashed line) approach their NGFP value as $k \to \infty$, while in the IR limit $k \to 0$, they scale as the mass dimension of their dimensionful counterparts. In the case of $\gc$, this scaling receives logarithmic corrections due to graviton fluctuations. Since $[\GN] = [\GC] = -2$, this logarithmic running explains why $g(k)$ and $\gc(k)$ do not run parallel in the IR. Fixing the transition scale $k_0 = m_\text{Pl}$, the logarithm can be subtracted and the constant part of the Wilson coefficient extracted. This is shown in the right panel: the purple line shows the running of the ratio $\gc(k)/g(k)$, which in the IR follows the scaling~\eqref{eq:ratio_IR}, with $b=0.5358$ for the UV-complete trajectory; subtracting the logarithmic running, one is left with a numerical fit (green dashed line) which, based on Eqs.~\eqref{eq:ratio_IR} and~\eqref{eq:alphaIR}, $\gc(k)/g(k)$ to uniquely determine the Wilson coefficient $\GC$ (solid yellow line) via the intersection between the constant part of the ratio $\gc(k)/g(k)$ and the line $k=k_0= m_{\text{Pl}}\Leftrightarrow\tau=0$ (dotted black line).}
    \label{fig:c2andg}
\end{figure}

Once the right initial condition is identified, the calculation of the Wilson coefficient $\GC$ requires the numerical integration of the coupled beta functions and a controlled treatment of the IR logarithmic behavior.
The machinery described in Sect.~\ref{s:RGflowanalysis} to subtract the logarithmic running can now be applied to this end. As evident from the left panel of Fig.~\ref{fig:c2andg}, $\gc$ runs logarithmically according to Eq.~\eqref{eq:IR_scaling}, so that the two couplings do not run parallel in the IR. 
Having fixed $k_0=m_{\text{Pl}}$ and $g_\text{IR}=1$,
we can compare the ansatz in Eq.~\eqref{eq:ratio_IR} with the numerical flow to extract the ($k_0$-independent) value for the slope, which is $b=0.5358$, and subtract the logarithm according to the prescription~\eqref{eq:prescription_G_N}. Next, Eq.~\eqref{eq:alphaIR} is employed to fix the Wilson coefficient for the Weyl-squared term,
\be \label{eq:Wilson_Coeff}
\GC=0.5092 \, m_\text{Pl}^{-2}\quad \Rightarrow \quad m_2 = 1.4013 \, m_\text{Pl} \,.
\ee
The procedure to extract this value is depicted in the right panel of Fig.~\ref{fig:c2andg}. 

Finally, let us comment on how the result given in Eq.~\eqref{eq:Wilson_Coeff} --- in particular, the positivity of the Weyl-squared Wilson coefficient --- changes if one considers a different prescription for the subtraction of the logarithmic running, involving a QG scale other than the Planck scale~\cite{DelPorro:2025fiu}. Setting $k_0= \xi \, m_{\rm Pl}$, our result remains qualitative unaltered if $\xi \gtrsim \mathcal{O}(1)$. That is, $\GC$ remains positive for
\be
\xi \geq 0.15 \,.
\ee
When $\xi$ approaches this lower bound, the predicted value of $\GC$ lowers to $0$, giving a higher and higher value of $m_2$. In this limit, any beyond-GR solution becomes effectively indistinguishable from a Schwarzschild black hole. Further decreasing the value of $\xi$ beyond the critical value $\xi\simeq 0.15$ eventually renders $\GC$ negative, making $m_2^2$ cross a discontinuity beyond which it becomes negative. This case would be problematic for our analysis: a negative $m_2^2$ would imply a different regularization for the modified inverse propagator in the Wetterich equation, since the sign of the regulator has to match that of the kinetic term in order to avoid unphysical poles~\cite{Percacci:2017fkn,Reuter:2019byg}. This would thereby alter the beta functions of~\cite{Knorr:2021slg}, invalidating our analysis. We stress, however, that in a fundamental realization of ASQG, the Planck scale --- corresponding to $\xi=\mathcal{O}(1)$ ---- is the only available scale of the system. Hence, within the natural assumption $k_0= \mathcal{O}( m_{\rm Pl})$, our result --- and the qualitative picture arising from it (see Sect.~\ref{sec:BH-from-ASEW}) --- remains solid.

In the following section, we shall see how this prediction for the Weyl-squared Wilson coefficient constrains the phase diagram of Einstein-Weyl gravity, described in Sect.~\ref{sect:PDconstr}.

\section{Mapping out GLOBs from an asymptotically safe UV completion}\label{sec:BH-from-ASEW}

With the value of $\GC$ fixed by ASQG, cf. Eq.~\eqref{eq:Wilson_Coeff}, we can now constrain the classical phase diagram (Fig.~\ref{fig:pd-review}) introduced in Sect.~\ref{sect:PDconstr}.

\begin{figure}[t!]
    \centering
    \begin{subfigure}[b]{0.48\linewidth}
        \centering
        \includegraphics[width=\linewidth]{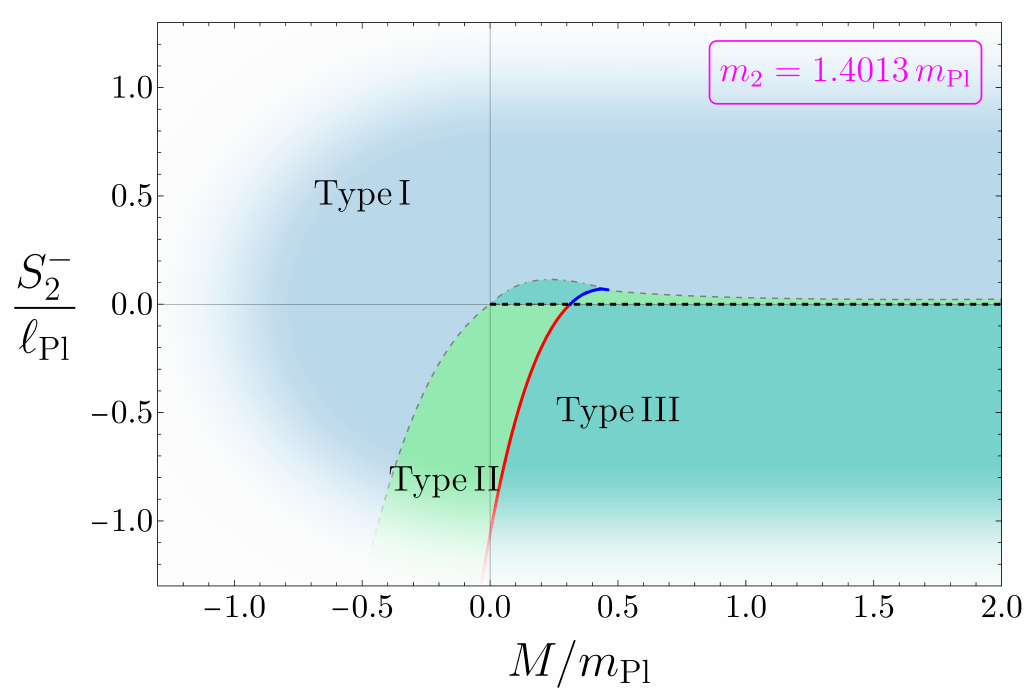}
    \end{subfigure}
    \hfill
    \begin{subfigure}[b]{0.48\linewidth}
        \centering
        \includegraphics[width=\linewidth]{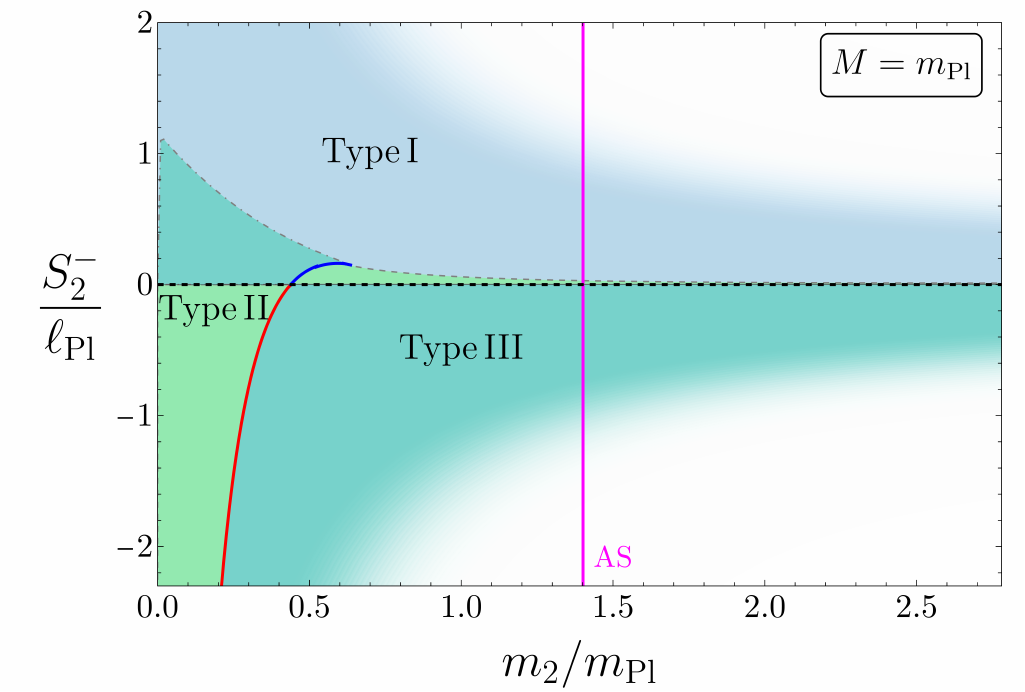}
    \end{subfigure}
    \vskip\baselineskip
    \begin{subfigure}[b]{0.48\linewidth}
        \centering
        \includegraphics[width=\linewidth]{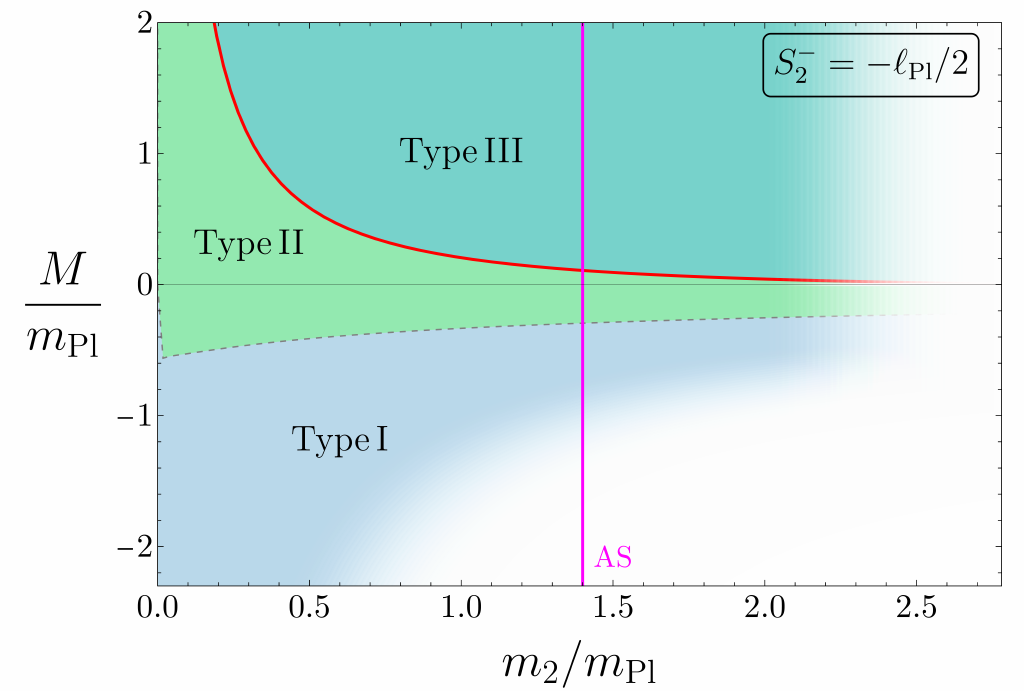}
    \end{subfigure}
    \hfill
    \begin{subfigure}[b]{0.48\linewidth}
        \centering
        \includegraphics[width=\linewidth]{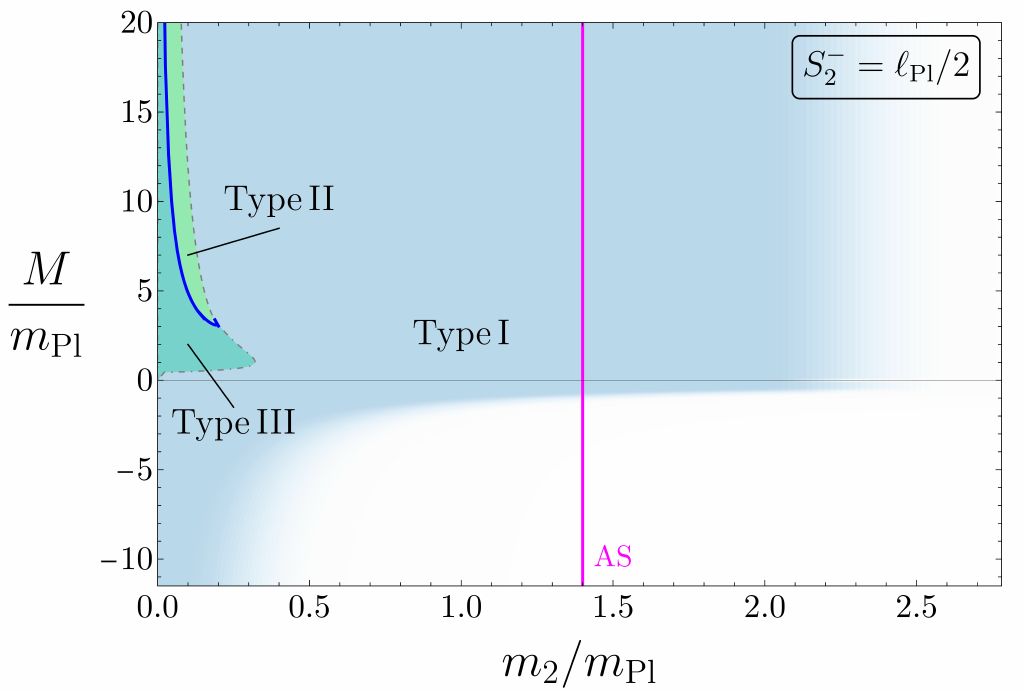}
    \end{subfigure}

    \caption{Different projections of the 3D GLOB phase diagram displaying the constraints from the requirement of UV completion within ASQG (magenta lines). The consistency conditions for the validity of the asymptotic analysis of~\cite{Silveravalle:2022wij}, which we derived in Sect.~\ref{sec:RegionofValidity}, are also accounted for: in all plots, the white areas are regions where the analysis of~\cite{Silveravalle:2022wij} is not applicable, whereas the colorful ones display the possible solutions in the region of validity, following the same color coding as in Fig.~\ref{fig:pd-review}~\cite{Silveravalle:2022wij}. The top-left panel represents a rescaled version of the classical diagram in Fig.~\ref{fig:pd-review}, obtained by setting $m_2$ to its value in ASQG: $m_2 = 1.4013 \ m_\text{Pl}$. The top-right figure represents a projection of the phase diagram onto the $\{ m_2/m_{\text{Pl}},\,S_2^-/l_{\text{Pl}},  \}$ plane at fixed ADM mass $M = m_\text{Pl}$. Here, our prediction for $m_2$ is represented by the magenta vertical line. Similarly, the two figures below depict the phase diagram in the $\{m_2/m_{\text{Pl}}, \, M/m_{\text{Pl}} \}$ projection for the two different values of the Yukawa charge, $S_2^-=\pm \ell_{\rm Pl}/2$. Again, the ASQG prediction for $m_2$ is shown as magenta vertical lines. }
    \label{fig:updated-diagram}
\end{figure}

Fig.~\ref{fig:pd-review} displays the phase diagram in units of $m_2$. Varying $m_2$ is tantamount to changing the scale on the axes. To see this more clearly, it is useful to use Planck units, so that the full phase diagram is a 3D plot given in terms of the dimensionless parameters $\{ S_2^-/l_{\text{Pl}}, \, M/m_{\text{Pl}}, \, m_2/m_{\text{Pl}}\}$. Any bi-dimensional $\{m_2={\rm const.}\}$-slice of this 3D plot has the same contours as Fig.~\ref{fig:pd-review}, with the scale of the axes fixed by the numerical value of~$m_2$. A prediction for $m_2$ from QG thus selects one single slice of the full 3D phase diagram, and hence constrains the spectrum of solutions of the theory.

Our computation, which determines the value of $m_2$ by requiring the theory to be UV complete, cf. Eq.~\eqref{eq:Wilson_Coeff}, thus selects the slice of the full 3D phase diagram predicted by ASQG. This is shown in the top-left panel of Fig.~\ref{fig:updated-diagram}. Moreover, as depicted in the other three panels of Fig.~\ref{fig:updated-diagram}, ASQG restricts the types of allowed GLOBs for any given ADM mass $M/m_\text{Pl}$ and Yukawa charge $S_2^-/l_\text{Pl}$. Accounting for the considerations we made in Sect.~\ref{sec:RegionofValidity}, in the new, ASQG-induced phase diagram, Fig.~\ref{fig:updated-diagram}, we have also blurred out the regions that are not consistent with the weak-field expansion performed in~\cite{Silveravalle:2022wij}. More explicitly, the white regions are those where the analysis of~\cite{Silveravalle:2022wij} is not applicable. In particular, their location and extension generally depend on the Weyl-squared Wilson coefficient $\GC$. 

Given that in our case $m_2$ is of Planckian order, the resulting set of possible static, spherically symmetric solutions of Einstein-Weyl gravity is restricted to the region around the Schwarzschild line where $S_2^-$ still remains of the order of the Planck length. All the GLOBs described in Sect.~\ref{sect:setup} are possible in principle, but, as apparent from Fig.~\ref{fig:updated-diagram}, some of them are disfavored by ASQG. To see this, let us focus on the orthogonal projections of the full phase diagram, namely on constant-$M$ and constant-$S_2^-$ slices of the latter. A sample of these slices is shown in the top-right and bottom panels of Fig.~\ref{fig:updated-diagram}. The prediction of ASQG for $m_2$ (vertical magenta lines) restricts the types of GLOBs for any given ADM mass and Yukawa charge. Fig.~\ref{fig:updated-diagram} shows this explicitly for Planckian-sized $M$ and $S_2^-$, highlighting two main features:
\begin{itemize}
    \item Type-II solutions (attractive naked singularities) are generically disfavored, in particular in the limit of large masses;
    \item Type-III solutions (non-symmetric wormholes) are highly favored by $S_2^-<0$, and repulsive naked singularities by $S_2^->0$.
\end{itemize}
As expected, the QG-based calculation that led us to Eq.~\eqref{eq:Wilson_Coeff} has guided us in the exclusion of some spherically symmetric solutions in the Einstein-Weyl truncation, which are in principle allowed at the classical level. This is a clear example of how first principles (such as the existence of a UV completion for gravity) can have a direct impact on black hole physics.

\section{Discussion and conclusions} \label{sec:conclusions}

In this work, we presented a proof of concept on how quantum gravity (QG) can constrain the set of admissible low-energy effective field theories (EFTs) and how these can, in turn, filter out the landscape of possible black hole spacetimes and all their alternatives. For the purpose of this and future work, we chose to collectively refer to these as ``gravitationally localized objects'' (GLOBs). 

Concretely, we focused on the Einstein-Weyl truncation, for which the landscape of possible (classical) vacuum, static and spherically symmetric solutions has been classified~\cite{Silveravalle:2022wij}. This classification is apparent through a ``phase diagram''~\cite{Silveravalle:2022wij} displaying these solutions as a function of interaction couplings and integration constants, such as the ADM mass. The crucial observation is that not all the classically-allowed solutions can come from QG: each of them --- at least in the large-mass limit --- ought to stem from a principle of least action from a specific EFT~\cite{Knorr:2022kqp}, and not all EFTs can come from an ultraviolet (UV) completion of gravity~\cite{Palti:2019pca}. Our work shows how one specific UV completion of gravity --- asymptotically safe quantum gravity (ASQG) --- imposes constraints on the GLOB phase diagram.

In classical general relativity (GR), the vacuum, static and spherically symmetric solutions are uniquely determined by a single-parameter family, namely, the Schwarzschild geometry with ADM mass $M$. The addition of higher-derivative operators --- such as $C^2$ --- in the gravitational Lagrangian can drastically change this structure, giving rise to a richer class of geometries, wherein Schwarzschild black holes form a measure-zero subset. In the weak-field limit, they can be classified in terms of the Weyl-square coupling $\GC=m_2^{-2}$ and two asymptotic integration constants: the ADM mass $M$ and the ``Yukawa charge'' $S_2^-$. The resulting 3D phase diagram includes non-Schwarzschild black holes, wormholes and two types of naked singularities. Interestingly, this 3D space enjoys a special property: each $\{m_2=\text{const.} \}$-slice looks the same, i.e., on each $\{ M/m_{\text{Pl}}, S_2^-/l_{\text{Pl}}\}$-projection, $m_2$ solely sets the scale of the axes~\cite{Silveravalle:2022wij}. Following~\cite{Silveravalle:2022wij}, and as already mentioned, we refer to the $\{ S_2^-/l_{\text{Pl}}, \, M/m_{\text{Pl}}, \, m_2/m_{\text{Pl}}\}$ plot as a phase diagram, where the different types of GLOBs play the role of the thermodynamical phases. Whether the analogy could be pushed further by treating these different solutions as actual thermodynamical phases remains unclear. Undoubtedly, answering this question and understanding the types of transformations that would enable transitions between phases in the diagram would be a valuable direction for future work.

Our work focused on two aspects: first, we revised the classical phase diagram of~\cite{Silveravalle:2022wij}, deriving conditions on its domain of applicability, and second, we derived a prediction for the Weyl-squared Wilson coefficient from ASQG, in order to constrain the range of possible GLOB solutions. To address the first point, in Sect.~\ref{sec:RegionofValidity}, we revised the derivation in~\cite{Silveravalle:2022wij}, and obtained conditions for the validity of the asymptotic analysis. These constraints limit the region of the parameter space where the derivation of~\cite{Silveravalle:2022wij} can be trusted and where the phase diagram effectively makes sense. Interestingly, the updated diagram is reliable for those geometries that deviate from the Schwarzschild line by $S_2^- \simeq m_2^{-1}$. In the second part, we derived a quantitative prediction for the Wilson coefficient $\GC$, imposing UV completeness of the theory within the asymptotic safety scenario. Studying the renormalization group (RG) of the Einstein-Weyl truncation, i.e., the flow projected onto the $\{ G_N, \GC \}$ theory space, we found that the UV completion is given by a non-Gaussian fixed point (NGFP) --- the Reuter fixed point underlying the ASQG program --- coming with a single relevant direction. As a consequence, the unique UV-complete RG trajectory identifies a single low-energy EFT and hence returns a prediction for the Weyl-squared Wilson coefficient $\GC$. Due to the logarithmic running of the Weyl-squared coupling, as is well-known in the EFT literature, the definition of the corresponding Wilson coefficient is ambiguous. To extract a prediction for $\GC$, we followed the prescription introduced in~\cite{Basile:2021krr,Knorr:2024yiu} and subtracted the logarithmic running $\log(k/k_0)$, with the QG scale $k_0$ set to be the Planck mass. With this prescription, we obtained a prediction for the Weyl-squared Wilson coefficient: $m_2=\sqrt{\GC} \simeq 1.4  \, m_{\rm Pl}$.

A few comments are in order. While fixing $k_0$ to the Planck scale appears natural, a rigorous definition of Wilson coefficients in the presence of logarithmic running remains a problem to be addressed in the future. At the same time, as stressed in Sect.~\ref{sec:prediction}, a different choice for $k_0$ does not have dramatic consequences on our conclusions. We explicitly checked that the qualitative picture remains unaltered if $k_0$ is chosen to be $k_0 \gtrsim 10^{-1} m_{\rm Pl}$. This makes the qualitative outcome of our work robust, since in a fundamental realization of ASQG, the Planck mass is expected to set the scale of QG. A rigorous and less ambiguous approach might be to include these logarithmic dependencies at the level of the effective action. It is well-known indeed that the quantum effective action encodes the physical running of the couplings as form factors~\cite{Knorr:2019atm,Knorr:2022dsx}, and logarithmic form factors are expected on general grounds. These form factors could still modify the EOMs together with their solutions. Characterizing the corresponding global solutions is expected to be much more involved than in the case of local effective actions, but, on the positive side, the definition of the Wilson coefficient would avoid the aforementioned ambiguities. This investigation might be a natural future perspective of this work.

The predicted value of the Weyl-squared coefficient allows us to draw some conclusions regarding the landscape of possible GLOBs compatible with an asymptotically safe UV completion.

The conditions we derived on validity of the asymptotic analysis of~\cite{Silveravalle:2022wij}, together with our prediction that $m_2/m_\text{Pl}\sim\mathcal{O}(1)$ imply that, by consistency, $S_2^-/\ell_{\rm Pl}\sim\mathcal{O}(1) $. We stress that the latter is not a physical constraint, rather a computational limitation dictated by the weak-field expansion in~\cite{Silveravalle:2022wij}. This limitation only allows us to consistently discuss GLOBs with Planckian-sized deviations from the Schwarzschild line, i.e., those geometries closely resembling Schwarzschild black holes in the asymptotic region. Limiting ourselves to these geometries, our calculation shows that attractive naked singularities are highly disfavored by ASQG, whereas wormholes and repulsive naked singularities remain two possible outcomes.

We stress that all these solutions present a singularity. However, this is not in contrast with the common expectation that a quantum description of gravity should cure singularities. More precisely, it is not (necessarily) a problem of ASQG, rather an expected outcome of the chosen approximation of the gravitational dynamics: our calculation can only restrict the landscape of GLOBs among the classically available ones within the chosen truncation, and both Einstein-Weyl and quadratic gravity are not expected to resolve singularities~\cite{Holdom:2002xy,Borissova:2020knn,Borissova:2023kzq}. A general treatment would require applying our analysis to include a higher number of operators in the gravitational sector. Together with the inclusion of the other quadratic terms, $\mathcal{R}^2$, this would mean considering cubic operators, $\mathcal{R}^3$. This set of operators crucially includes the Goroff-Sagnotti counterterm~\cite{Goroff:1985th,Goroff:1985sz,vandeVen:1991gw}, which has been shown to be asymptotically safe~\cite{Gies:2016con,Baldazzi:2023pep}. This kind of extension has also been a recent object of study, since it has been suggested that some black hole solutions may be sensitive to higher-derivative corrections already at the level of their horizon \cite{Horowitz:2023xyl}, see also~\cite{DelPorro:2025fiu}. Notably, $\mathcal{R}^3$ is also the first set of operators possibly leading to non-singular GLOBs~\cite{Holdom:2002xy,Borissova:2023kzq,Bueno:2024eig}. Extending our analysis beyond quadratic order, possibly including essential operators only~\cite{Baldazzi:2021orb,Baldazzi:2021ydj,Baldazzi:2023pep}, is also desirable to avoid unphysical features related to the spin-2 ghost pole. In a truncation of the effective action, this pole can generally be fictitious~\cite{Platania:2020knd}, and indeed the non-perturbative dressed propagator in ASQG seems to be ghost-free~\cite{Knorr:2021niv,Fehre:2021eob}. However, a complete classification of all possible classical solutions of such a theory, even in the static, spherically symmetric case, is still missing. We defer this important step to future work.

All in all, our findings underscore the important role of QG in constraining black hole physics. The necessity of a UV completion is not merely a high-energy feature; it is a predictive principle that narrows down the set of viable EFTs and hence the possible GLOBs. We hope that this work, beyond its intrinsic relevance for the comprehension of GLOBs in asymptotically safe Einstein-Weyl gravity, would serve as a proof of concept and would stimulate further explorations in similar directions.

\acknowledgments
The authors would like to thank B. Knorr for providing the beta functions of~\cite{Knorr:2021slg} and I. Basile and S. Liberati for discussions.  The research of A.P. and F.D.P. is supported by a research grant (VIL60819) from VILLUM FONDEN. The Center of Gravity is a Center of Excellence funded by the Danish National Research Foundation under grant No. 184. The research of S.S. is funded by the European Union - NextGenerationEU under the PRIN MUR 2022 project. 20224JR28W `Charting unexplored avenues in Dark Matter'.

\printbibliography{}

@article{Fehre:2021eob,
    author = "Fehre, Jannik and Litim, Daniel F. and Pawlowski, Jan M. and Reichert, Manuel",
    title = "{Lorentzian Quantum Gravity and the Graviton Spectral Function}",
    eprint = "2111.13232",
    archivePrefix = "arXiv",
    primaryClass = "hep-th",
    doi = "10.1103/PhysRevLett.130.081501",
    journal = "Phys. Rev. Lett.",
    volume = "130",
    number = "8",
    pages = "081501",
    year = "2023"
}

@article{Biemans:2017zca,
    author = "Biemans, Jorn and Platania, Alessia and Saueressig, Frank",
    title = "{Renormalization group fixed points of foliated gravity-matter systems}",
    eprint = "1702.06539",
    archivePrefix = "arXiv",
    primaryClass = "hep-th",
    doi = "10.1007/JHEP05(2017)093",
    journal = "JHEP",
    volume = "05",
    pages = "093",
    year = "2017"
}

@article{Pawlowski:2025etp,
    author = "Pawlowski, Jan M. and Reichert, Manuel and Wessely, Jonas",
    title = "{Self-consistent graviton spectral function in Lorentzian quantum gravity}",
    eprint = "2507.22169",
    archivePrefix = "arXiv",
    primaryClass = "hep-th",
    month = "7",
    year = "2025"
}

@article{Banerjee:2022xvi,
    author = "Banerjee, Rudrajit and Niedermaier, Max",
    title = "{The spatial Functional Renormalization Group and Hadamard states on cosmological spacetimes}",
    eprint = "2201.02575",
    archivePrefix = "arXiv",
    primaryClass = "hep-th",
    doi = "10.1016/j.nuclphysb.2022.115814",
    journal = "Nucl. Phys. B",
    volume = "980",
    pages = "115814",
    year = "2022"
}

@article{DAngelo:2022vsh,
    author = "D'Angelo, Edoardo and Drago, Nicol\`o and Pinamonti, Nicola and Rejzner, Kasia",
    title = "{An Algebraic QFT Approach to the Wetterich Equation on Lorentzian Manifolds}",
    eprint = "2202.07580",
    archivePrefix = "arXiv",
    primaryClass = "math-ph",
    doi = "10.1007/s00023-023-01348-4",
    journal = "Annales Henri Poincare",
    volume = "25",
    number = "4",
    pages = "2295--2352",
    year = "2024"
}

@unpublished{DAngelo:2023tis,
    author = "D'Angelo, Edoardo and Rejzner, Kasia",
    title = "{A Lorentzian renormalisation group equation for gauge theories}",
    eprint = "2303.01479",
    archivePrefix = "arXiv",
    primaryClass = "math-ph",
    month = "3",
    year = "2023",
    note = "{a}rXiv Preprint"
}

@article{Manrique:2011jc,
    author = "Manrique, Elisa and Rechenberger, Stefan and Saueressig, Frank",
    title = "{Asymptotically Safe Lorentzian Gravity}",
    eprint = "1102.5012",
    archivePrefix = "arXiv",
    primaryClass = "hep-th",
    reportNumber = "MZ-TH-11-02",
    doi = "10.1103/PhysRevLett.106.251302",
    journal = "Phys. Rev. Lett.",
    volume = "106",
    pages = "251302",
    year = "2011"
}

@article{Biemans:2016rvp,
    author = "Biemans, Jorn and Platania, Alessia and Saueressig, Frank",
    title = "{Quantum gravity on foliated spacetimes: Asymptotically safe and sound}",
    eprint = "1609.04813",
    archivePrefix = "arXiv",
    primaryClass = "hep-th",
    doi = "10.1103/PhysRevD.95.086013",
    journal = "Phys. Rev. D",
    volume = "95",
    number = "8",
    pages = "086013",
    year = "2017"
}

@article{Knorr:2018fdu,
    author = "Knorr, Benjamin",
    title = "{Lorentz symmetry is relevant}",
    eprint = "1810.07971",
    archivePrefix = "arXiv",
    primaryClass = "hep-th",
    doi = "10.1016/j.physletb.2019.01.070",
    journal = "Phys. Lett. B",
    volume = "792",
    pages = "142--148",
    year = "2019"
}

@article{Eichhorn:2019ybe,
    author = "Eichhorn, Astrid and Platania, Alessia and Schiffer, Marc",
    title = "{Lorentz invariance violations in the interplay of quantum gravity with matter}",
    eprint = "1911.10066",
    archivePrefix = "arXiv",
    primaryClass = "hep-th",
    doi = "10.1103/PhysRevD.102.026007",
    journal = "Phys. Rev. D",
    volume = "102",
    number = "2",
    pages = "026007",
    year = "2020"
}

@article{Christiansen:2017bsy,
    author = "Christiansen, Nicolai and Falls, Kevin and Pawlowski, Jan M. and Reichert, Manuel",
    title = "{Curvature dependence of quantum gravity}",
    eprint = "1711.09259",
    archivePrefix = "arXiv",
    primaryClass = "hep-th",
    doi = "10.1103/PhysRevD.97.046007",
    journal = "Phys. Rev. D",
    volume = "97",
    number = "4",
    pages = "046007",
    year = "2018"
}

@article{Knorr:2017fus,
    author = "Knorr, Benjamin and Lippoldt, Stefan",
    title = "{Correlation functions on a curved background}",
    eprint = "1707.01397",
    archivePrefix = "arXiv",
    primaryClass = "hep-th",
    doi = "10.1103/PhysRevD.96.065020",
    journal = "Phys. Rev. D",
    volume = "96",
    number = "6",
    pages = "065020",
    year = "2017"
}

@article{Borissova:2025frj,
    author = "Borissova, Johanna and Dittrich, Bianca and Eichhorn, Astrid and Schiffer, Marc",
    title = "{Renormalization group flows in area-metric gravity}",
    eprint = "2507.02034",
    archivePrefix = "arXiv",
    primaryClass = "gr-qc",
    month = "7",
    year = "2025"
}

@article{Nink:2014yya,
    author = "Nink, Andreas",
    title = "{Field Parametrization Dependence in Asymptotically Safe Quantum Gravity}",
    eprint = "1410.7816",
    archivePrefix = "arXiv",
    primaryClass = "hep-th",
    doi = "10.1103/PhysRevD.91.044030",
    journal = "Phys. Rev. D",
    volume = "91",
    number = "4",
    pages = "044030",
    year = "2015"
}

@article{DeBrito:2018hur,
    author = "De Brito, Gustavo P. and Ohta, Nobuyoshi and Pereira, Antonio D. and Tomaz, Anderson A. and Yamada, Masatoshi",
    title = "{Asymptotic safety and field parametrization dependence in the $f(R)$ truncation}",
    eprint = "1805.09656",
    archivePrefix = "arXiv",
    primaryClass = "hep-th",
    doi = "10.1103/PhysRevD.98.026027",
    journal = "Phys. Rev. D",
    volume = "98",
    number = "2",
    pages = "026027",
    year = "2018"
}

@article{Kluth:2022vnq,
    author = "Kluth, Yannick and Litim, Daniel F.",
    title = "{Functional renormalization for f(R{\ensuremath{\mu}}{\ensuremath{\nu}}{\ensuremath{\rho}}{\ensuremath{\sigma}}) quantum gravity}",
    eprint = "2202.10436",
    archivePrefix = "arXiv",
    primaryClass = "hep-th",
    doi = "10.1103/PhysRevD.106.106022",
    journal = "Phys. Rev. D",
    volume = "106",
    number = "10",
    pages = "106022",
    year = "2022"
}

@article{Christiansen:2017cxa,
    author = "Christiansen, Nicolai and Litim, Daniel F. and Pawlowski, Jan M. and Reichert, Manuel",
    title = "{Asymptotic safety of gravity with matter}",
    eprint = "1710.04669",
    archivePrefix = "arXiv",
    primaryClass = "hep-th",
    doi = "10.1103/PhysRevD.97.106012",
    journal = "Phys. Rev. D",
    volume = "97",
    number = "10",
    pages = "106012",
    year = "2018"
}

@article{Falls:2014tra,
    author = "Falls, Kevin and Litim, Daniel F. and Nikolakopoulos, Konstantinos and Rahmede, Christoph",
    title = "{Further evidence for asymptotic safety of quantum gravity}",
    eprint = "1410.4815",
    archivePrefix = "arXiv",
    primaryClass = "hep-th",
    reportNumber = "DO-TH-14-26, KA-TP-2014-30",
    doi = "10.1103/PhysRevD.93.104022",
    journal = "Phys. Rev. D",
    volume = "93",
    number = "10",
    pages = "104022",
    year = "2016"
}

@article{Falls:2017lst,
    author = "Falls, Kevin and King, Callum R. and Litim, Daniel F. and Nikolakopoulos, Kostas and Rahmede, Christoph",
    title = "{Asymptotic safety of quantum gravity beyond Ricci scalars}",
    eprint = "1801.00162",
    archivePrefix = "arXiv",
    primaryClass = "hep-th",
    doi = "10.1103/PhysRevD.97.086006",
    journal = "Phys. Rev. D",
    volume = "97",
    number = "8",
    pages = "086006",
    year = "2018"
}

@article{Kluth:2020bdv,
    author = "Kluth, Yannick and Litim, Daniel F.",
    title = "{Fixed points of quantum gravity and the dimensionality of the UV critical surface}",
    eprint = "2008.09181",
    archivePrefix = "arXiv",
    primaryClass = "hep-th",
    doi = "10.1103/PhysRevD.108.026005",
    journal = "Phys. Rev. D",
    volume = "108",
    number = "2",
    pages = "026005",
    year = "2023"
}

@article{Palti:2019pca,
    author = "Palti, Eran",
    title = "{The Swampland: Introduction and Review}",
    eprint = "1903.06239",
    archivePrefix = "arXiv",
    primaryClass = "hep-th",
    reportNumber = "MPP-2019-53",
    doi = "10.1002/prop.201900037",
    journal = "Fortsch. Phys.",
    volume = "67",
    number = "6",
    pages = "1900037",
    year = "2019"
}

@article{Dupuis:2020fhh,
    author = "Dupuis, N. and Canet, L. and Eichhorn, A. and Metzner, W. and Pawlowski, J. M. and Tissier, M. and Wschebor, N.",
    title = "{The nonperturbative functional renormalization group and its applications}",
    eprint = "2006.04853",
    archivePrefix = "arXiv",
    primaryClass = "cond-mat.stat-mech",
    doi = "10.1016/j.physrep.2021.01.001",
    journal = "Phys. Rept.",
    volume = "910",
    pages = "1--114",
    year = "2021"
}

@article{Daas:2022iid,
    author = "Daas, Jesse and Kuijpers, Kolja and Saueressig, Frank and Wondrak, Michael F. and Falcke, Heino",
    title = "{Probing quadratic gravity with the Event Horizon Telescope}",
    eprint = "2204.08480",
    archivePrefix = "arXiv",
    primaryClass = "gr-qc",
    doi = "10.1051/0004-6361/202244080",
    journal = "Astron. Astrophys.",
    volume = "673",
    pages = "A53",
    year = "2023"
}

@article{Nelson:2010ig,
    author = "Nelson, William",
    title = "{Static Solutions for 4th order gravity}",
    eprint = "1010.3986",
    archivePrefix = "arXiv",
    primaryClass = "gr-qc",
    doi = "10.1103/PhysRevD.82.104026",
    journal = "Phys. Rev. D",
    volume = "82",
    pages = "104026",
    year = "2010"
}

@inbook{Knorr:2022dsx,
    author = "Knorr, Benjamin and Ripken, Chris and Saueressig, Frank",
    title = "{Form Factors in Asymptotically Safe Quantum Gravity}",
    eprint = "2210.16072",
    archivePrefix = "arXiv",
    primaryClass = "hep-th",
    reportNumber = "NORDITA 2022-075",
    doi = "10.1007/978-981-19-3079-9_21-1",
    bookTitle="Handbook of Quantum Gravity",
    year="2023",
    publisher="Springer Nature Singapore",
    address="Singapore",
    pages="1--49"
}

@inbook{Eichhorn:2022gku,
    author = "Eichhorn, Astrid and Schiffer, Marc",
    title = "Asymptotic Safety of Gravity with Matter",
    eprint = "2212.07456",
    archivePrefix = "arXiv",
    primaryClass = "hep-th",
    bookTitle="Handbook of Quantum Gravity",
    year="2023",
    publisher="Springer Nature Singapore",
    address="Singapore",
    pages="1--87"
}

@inbook{Morris:2022btf,
    author = "Morris, Tim R. and Stulga, Dalius",
    title = "{The Functional f(R) Approximation}",
    eprint = "2210.11356",
    archivePrefix = "arXiv",
    primaryClass = "hep-th",
    doi = "10.1007/978-981-19-3079-9_19-1",
    bookTitle="Handbook of Quantum Gravity",
    year="2023",
    publisher="Springer Nature Singapore",
    address="Singapore",
    pages="1--33"
}

@inbook{Martini:2022sll,
    author = "Martini, Riccardo and Vacca, Gian Paolo and Zanusso, Omar",
    title = "{Perturbative Approaches to Nonperturbative Quantum Gravity}",
    booktitle = "{Handbook of Quantum Gravity}",
    eprint = "2210.13910",
    archivePrefix = "arXiv",
    primaryClass = "hep-th",
    doi = "10.1007/978-981-19-3079-9_25-1",
    year="2023",
    publisher="Springer Nature Singapore",
    address="Singapore",
    pages="1--46"
}

@inbook{Wetterich:2022ncl,
    author = "Wetterich, C.",
    title = "{Quantum Gravity and Scale Symmetry in Cosmology}",
    eprint = "2211.03596",
    archivePrefix = "arXiv",
    primaryClass = "gr-qc",
    doi = "10.1007/978-981-19-3079-9_26-1",
    bookTitle="Handbook of Quantum Gravity",
    year="2023",
    publisher="Springer Nature Singapore",
    address="Singapore",
    pages="1--68"
}

@inbook{Platania:2023srt,
    author = "Platania, Alessia",
    title = "{Black Holes in Asymptotically Safe Gravity}",
    eprint = "2302.04272",
    archivePrefix = "arXiv",
    primaryClass = "gr-qc",
    reportNumber = "NORDITA 2022-085",
    doi = "10.1007/978-981-19-3079-9_24-1",
    bookTitle="Handbook of Quantum Gravity",
    year="2023",
    publisher="Springer Nature Singapore",
    address="Singapore",
    pages="1--65"
}

@inbook{Saueressig:2023irs,
    author = "Saueressig, Frank",
    title = "{The Functional Renormalization Group in Quantum Gravity}",
    eprint = "2302.14152",
    archivePrefix = "arXiv",
    primaryClass = "hep-th",
    doi = "10.1007/978-981-19-3079-9_16-1",
    bookTitle="Handbook of Quantum Gravity",
    year="2023",
    publisher="Springer Nature Singapore",
    address="Singapore",
    pages="1--44"
}

@inbook{Pawlowski:2023gym,
    author = "Pawlowski, Jan M. and Reichert, Manuel",
    title = "{Quantum Gravity from Dynamical Metric Fluctuations}",
    eprint = "2309.10785",
    archivePrefix = "arXiv",
    primaryClass = "hep-th",
    bookTitle="Handbook of Quantum Gravity",
    year="2023",
    publisher="Springer Nature Singapore",
    address="Singapore",
    pages="1--70"
}

@article{Carballo-Rubio:2025fnc,
    author = "Carballo-Rubio, Ra\'ul and others",
    title = "{Towards a non-singular paradigm of black hole physics}",
    eprint = "2501.05505",
    archivePrefix = "arXiv",
    primaryClass = "gr-qc",
    doi = "10.1088/1475-7516/2025/05/003",
    journal = "JCAP",
    volume = "05",
    pages = "003",
    year = "2025"
}

@article{Eichhorn:2024wba,
    author = "Eichhorn, Astrid and Schiffer, Marc and Pedersen, Andreas Odgaard",
    title = "{Application of positivity bounds in asymptotically safe gravity}",
    eprint = "2405.08862",
    archivePrefix = "arXiv",
    primaryClass = "hep-th",
    doi = "10.1140/epjc/s10052-025-14449-7",
    journal = "Eur. Phys. J. C",
    volume = "85",
    number = "7",
    pages = "733",
    year = "2025"
}

@article{Knorr:2022kqp,
    author = "Knorr, Benjamin and Platania, Alessia",
    title = "{Sifting quantum black holes through the principle of least action}",
    eprint = "2202.01216",
    archivePrefix = "arXiv",
    primaryClass = "hep-th",
    doi = "10.1103/PhysRevD.106.L021901",
    journal = "Phys. Rev. D",
    volume = "106",
    number = "2",
    pages = "L021901",
    year = "2022"
}

@article{Platania:2023uda,
    author = "Platania, Alessia and Redondo-Yuste, Jaime",
    title = "{Diverging black hole entropy from quantum infrared non-localities}",
    eprint = "2303.17621",
    archivePrefix = "arXiv",
    primaryClass = "hep-th",
    doi = "10.1016/j.physletb.2024.138993",
    journal = "Phys. Lett. B",
    volume = "857",
    pages = "138993",
    year = "2024"
}

@article{Platania:2025imw,
    author = "Platania, Alessia",
    title = "{Some thoughts about black holes in asymptotic safety}",
    doi = "10.1007/s10714-025-03390-5",
    journal = "Gen. Rel. Grav.",
    volume = "57",
    number = "3",
    pages = "58",
    year = "2025"
}

@article{Buoninfante:2024oyi,
    author = "Buoninfante, Luca and Di Filippo, Francesco and Kol\'a\v{r}, Ivan and Saueressig, Frank",
    title = "{Dust collapse and horizon formation in quadratic gravity}",
    eprint = "2410.05941",
    archivePrefix = "arXiv",
    primaryClass = "gr-qc",
    doi = "10.1088/1475-7516/2025/01/114",
    journal = "JCAP",
    volume = "01",
    pages = "114",
    year = "2025"
}

@article{Pawlowski:2023dda,
    author = {Pawlowski, Jan M. and Tr\"ankle, Jan},
    title = "{Effective action and black hole solutions in asymptotically safe quantum gravity}",
    eprint = "2309.17043",
    archivePrefix = "arXiv",
    primaryClass = "hep-th",
    doi = "10.1103/PhysRevD.110.086011",
    journal = "Phys. Rev. D",
    volume = "110",
    number = "8",
    pages = "086011",
    year = "2024"
}

@article{Held:2020kze,
    author = "Held, Aaron",
    title = "{Effective asymptotic safety and its predictive power: Gauge-Yukawa theories}",
    eprint = "2003.13642",
    archivePrefix = "arXiv",
    primaryClass = "hep-th",
    reportNumber = "Imperial/TP/2020/AH/02",
    doi = "10.3389/fphy.2020.00341",
    journal = "Front. in Phys.",
    volume = "8",
    pages = "341",
    year = "2020"
}

@article{Basile:2021krk,
    author = "Basile, Ivano and Platania, Alessia",
    title = "{String tension between de Sitter vacua and curvature corrections}",
    eprint = "2103.06276",
    archivePrefix = "arXiv",
    primaryClass = "hep-th",
    doi = "10.1103/PhysRevD.104.L121901",
    journal = "Phys. Rev. D",
    volume = "104",
    number = "12",
    pages = "L121901",
    year = "2021"
}

@article{Basile:2021euh,
    author = "Basile, Ivano and Platania, Alessia",
    title = "{Cosmological \ensuremath{\alpha}'-corrections from the functional renormalization group}",
    eprint = "2101.02226",
    archivePrefix = "arXiv",
    primaryClass = "hep-th",
    doi = "10.1007/JHEP06(2021)045",
    journal = "JHEP",
    volume = "06",
    pages = "045",
    year = "2021"
}

@article{deAlwis:2019aud,
    author = "de Alwis, Senarath and Eichhorn, Astrid and Held, Aaron and Pawlowski, Jan M. and Schiffer, Marc and Versteegen, Fleur",
    title = "{Asymptotic safety, string theory and the weak gravity conjecture}",
    eprint = "1907.07894",
    archivePrefix = "arXiv",
    primaryClass = "hep-th",
    doi = "10.1016/j.physletb.2019.134991",
    journal = "Phys. Lett. B",
    volume = "798",
    pages = "134991",
    year = "2019"
}

@article{Dai:2023tud,
    author = "Dai, Mingwei and Freeman, Walter and Laiho, Jack and Schiffer, Marc and Unmuth-Yockey, Judah",
    title = "{Improved algorithm for dynamical triangulations and simulations of finer lattices}",
    eprint = "2309.12257",
    archivePrefix = "arXiv",
    primaryClass = "hep-lat",
    reportNumber = "FERMILAB-PUB-23-558-T",
    doi = "10.1103/PhysRevD.109.034518",
    journal = "Phys. Rev. D",
    volume = "109",
    number = "3",
    pages = "034518",
    year = "2024"
}

@article{Basile:2025zjc,
    author = "Basile, Ivano and Knorr, Benjamin and Platania, Alessia and Schiffer, Marc",
    title = "{Asymptotic safety, quantum gravity, and the swampland: a conceptual assessment}",
    eprint = "2502.12290",
    archivePrefix = "arXiv",
    primaryClass = "hep-th",
    reportNumber = "MPP-2025-37",
    month = "2",
    year = "2025"
}

@article{Loll:2019rdj,
    author = "Loll, R.",
    title = "{Quantum Gravity from Causal Dynamical Triangulations: A Review}",
    eprint = "1905.08669",
    archivePrefix = "arXiv",
    primaryClass = "hep-th",
    doi = "10.1088/1361-6382/ab57c7",
    journal = "Class. Quant. Grav.",
    volume = "37",
    number = "1",
    pages = "013002",
    year = "2020"
}

@article{Dymnikova:1992ux,
    author = "Dymnikova, I.",
    title = "{Vacuum nonsingular black hole}",
    doi = "10.1007/BF00760226",
    journal = "Gen. Rel. Grav.",
    volume = "24",
    pages = "235--242",
    year = "1992"
}

@article{Hayward:2005gi,
    author = "Hayward, Sean A.",
    title = "{Formation and evaporation of regular black holes}",
    eprint = "gr-qc/0506126",
    archivePrefix = "arXiv",
    doi = "10.1103/PhysRevLett.96.031103",
    journal = "Phys. Rev. Lett.",
    volume = "96",
    pages = "031103",
    year = "2006"
}

@inproceedings{Buoninfante:2024oxl,
    author = "Afshordi, Niayesh and others",
    editor = "Buoninfante, Luca and Carballo-Rubio, Ra\'ul and Cardoso, Vitor and Di Filippo, Francesco and Eichhorn, Astrid",
    title = "{Black Holes Inside and Out 2024: visions for the future of black hole physics}",
    eprint = "2410.14414",
    archivePrefix = "arXiv",
    primaryClass = "gr-qc",
    month = "10",
    year = "2024"
}

@article{Borissova:2022mgd,
    author = "Borissova, Johanna N. and Platania, Alessia",
    title = "{Formation and evaporation of quantum black holes from the decoupling mechanism in quantum gravity}",
    eprint = "2210.01138",
    archivePrefix = "arXiv",
    primaryClass = "gr-qc",
    reportNumber = "NORDITA 2022-069",
    doi = "10.1007/JHEP03(2023)046",
    journal = "JHEP",
    volume = "03",
    pages = "046",
    year = "2023"
}

@inbook{Bonanno:2024xne,
    author = "Bonanno, Alfio",
    title = "{Asymptotic Safety and Cosmology}",
    doi = "10.1007/978-981-19-3079-9_23-1",
    bookTitle="Handbook of Quantum Gravity",
    year="2023",
    publisher="Springer Nature Singapore",
    address="Singapore",
    pages="1--27"
}

@Inbook{Eichhorn:2022bgu,
author="Eichhorn, Astrid
and Held, Aaron",
editor="Bambi, Cosimo",
title="Black Holes in Asymptotically Safe Gravity and Beyond",
bookTitle="Regular Black Holes: Towards a New Paradigm of Gravitational Collapse",
year="2023",
publisher="Springer Nature Singapore",
address="Singapore",
pages="131--183",
isbn="978-981-99-1596-5",
doi="10.1007/978-981-99-1596-5_5",
url="https://doi.org/10.1007/978-981-99-1596-5_5"
}

@book{Polchinski:1998rq,
    author = "Polchinski, J.",
    title = "{String theory. Vol. 1: An introduction to the bosonic string}",
    doi = "10.1017/CBO9780511816079",
    isbn = "978-0-511-25227-3, 978-0-521-67227-6, 978-0-521-63303-1",
    publisher = "Cambridge University Press",
    series = "Cambridge Monographs on Mathematical Physics",
    month = "12",
    year = "2007"
}

@book{Polchinski:1998rr,
    author = "Polchinski, J.",
    title = "{String theory. Vol. 2: Superstring theory and beyond}",
    doi = "10.1017/CBO9780511618123",
    isbn = "978-0-511-25228-0, 978-0-521-63304-8, 978-0-521-67228-3",
    publisher = "Cambridge University Press",
    series = "Cambridge Monographs on Mathematical Physics",
    month = "12",
    year = "2007"
}

@article{Rovelli:1997yv,
    author = "Rovelli, Carlo",
    title = "{Loop quantum gravity}",
    eprint = "gr-qc/9710008",
    archivePrefix = "arXiv",
    doi = "10.12942/lrr-1998-1",
    journal = "Living Rev. Rel.",
    volume = "1",
    pages = "1",
    year = "1998"
}

@BOOK{Bambi2024-nm,
  title     = "Handbook of quantum gravity",
  editor    = "Bambi, Cosimo and Modesto, Leonardo and Shapiro, Ilya",
  publisher = "Springer Nature Singapore",
  year      =  2024,
  address   = "Singapore",
  copyright = "https://www.springernature.com/gp/researchers/text-and-data-mining"
}

@article{Bueno:2024eig,
    author = "Bueno, Pablo and Cano, Pablo A. and Hennigar, Robie A. and Murcia, {\'A}ngel J.",
    title = "{Dynamical Formation of Regular Black Holes}",
    eprint = "2412.02742",
    archivePrefix = "arXiv",
    primaryClass = "gr-qc",
    doi = "10.1103/PhysRevLett.134.181401",
    journal = "Phys. Rev. Lett.",
    volume = "134",
    number = "18",
    pages = "181401",
    year = "2025"
}

@article{Buoninfante:2024yth,
    author = "Buoninfante, Luca and others",
    title = "{Visions in Quantum Gravity}",
    eprint = "2412.08696",
    archivePrefix = "arXiv",
    primaryClass = "hep-th",
    journal={SciPost Phys. Comm. Rep.},
	pages={11},
	year={2025},
	publisher={SciPost},
	doi={10.21468/SciPostPhysCommRep.11}
}

@Article{Basile:2024oms,
	title={{Lectures in quantum gravity}},
	author={Ivano Basile and Luca Buoninfante and Francesco Di Filippo and Benjamin Knorr and Alessia Platania and Anna Tokareva},
	journal={SciPost Phys. Lect. Notes},
	pages={98},
	year={2025},
	publisher={SciPost},
	doi={10.21468/SciPostPhysLectNotes.98},
    eprint = "2412.08690",
    archivePrefix = "arXiv",
    primaryClass = "hep-th",
	url={https://scipost.org/10.21468/SciPostPhysLectNotes.98}
}

@article{Silveravalle:2022wij,
    author = "Silveravalle, Samuele and Zuccotti, Alessandro",
    title = "{Phase diagram of Einstein-Weyl gravity}",
    eprint = "2210.13877",
    archivePrefix = "arXiv",
    primaryClass = "gr-qc",
    doi = "10.1103/PhysRevD.107.064029",
    journal = "Phys. Rev. D",
    volume = "107",
    number = "6",
    pages = "064029",
    year = "2023"
}

@article{Lu:2017kzi,
    author = {L\"u, Hong and Perkins, A. and Pope, C. N. and Stelle, K. S.},
    title = "{Lichnerowicz Modes and Black Hole Families in Ricci Quadratic Gravity}",
    eprint = "1704.05493",
    archivePrefix = "arXiv",
    primaryClass = "hep-th",
    reportNumber = "IMPERIAL-TP-17-KSS-01, MI-TH-1750",
    doi = "10.1103/PhysRevD.96.046006",
    journal = "Phys. Rev. D",
    volume = "96",
    number = "4",
    pages = "046006",
    year = "2017"
}

@article{Held:2022abx,
    author = "Held, Aaron and Zhang, Jun",
    title = "{Instability of spherically symmetric black holes in quadratic gravity}",
    eprint = "2209.01867",
    archivePrefix = "arXiv",
    primaryClass = "gr-qc",
    reportNumber = "Imperial/TP/2022/AH/03",
    doi = "10.1103/PhysRevD.107.064060",
    journal = "Phys. Rev. D",
    volume = "107",
    number = "6",
    pages = "064060",
    year = "2023"
}

@article{Holdom:2002xy,
    author = "Holdom, Bob",
    title = "{On the fate of singularities and horizons in higher derivative gravity}",
    eprint = "hep-th/0206219",
    archivePrefix = "arXiv",
    reportNumber = "UTPT-02-09",
    doi = "10.1103/PhysRevD.66.084010",
    journal = "Phys. Rev. D",
    volume = "66",
    pages = "084010",
    year = "2002"
}

@article{Holdom:2016nek,
    author = "Holdom, Bob and Ren, Jing",
    title = "{Not quite a black hole}",
    eprint = "1612.04889",
    archivePrefix = "arXiv",
    primaryClass = "gr-qc",
    doi = "10.1103/PhysRevD.95.084034",
    journal = "Phys. Rev. D",
    volume = "95",
    number = "8",
    pages = "084034",
    year = "2017"
}

@article{Stelle:1977ry,
    author = "Stelle, K. S.",
    title = "{Classical Gravity with Higher Derivatives}",
    reportNumber = "Print-77-0417 (BRANDEIS)",
    doi = "10.1007/BF00760427",
    journal = "Gen. Rel. Grav.",
    volume = "9",
    pages = "353--371",
    year = "1978"
}

@article{Bonanno:2022ibv,
    author = "Bonanno, Alfio and Silveravalle, Samuele and Zuccotti, Alessandro",
    title = "{Nonsymmetric wormholes and localized big rip singularities in Einstein-Weyl gravity}",
    eprint = "2204.04966",
    archivePrefix = "arXiv",
    primaryClass = "gr-qc",
    doi = "10.1103/PhysRevD.105.124059",
    journal = "Phys. Rev. D",
    volume = "105",
    number = "12",
    pages = "124059",
    year = "2022"
}

@article{Lu:2015cqa,
    author = "Lu, H. and Perkins, A. and Pope, C. N. and Stelle, K. S.",
    title = "{Black Holes in Higher-Derivative Gravity}",
    eprint = "1502.01028",
    archivePrefix = "arXiv",
    primaryClass = "hep-th",
    reportNumber = "IMPERIAL-TP-15-KSS-01, MI-TH-1504, CAQS-1501",
    doi = "10.1103/PhysRevLett.114.171601",
    journal = "Phys. Rev. Lett.",
    volume = "114",
    number = "17",
    pages = "171601",
    year = "2015"
}

@article{Lu:2015psa,
    author = {L\"u, H. and Perkins, A. and Pope, C. N. and Stelle, K. S.},
    title = "{Spherically Symmetric Solutions in Higher-Derivative Gravity}",
    eprint = "1508.00010",
    archivePrefix = "arXiv",
    primaryClass = "hep-th",
    reportNumber = "IMPERIAL-TP-15-KSS-02, MI-TH-1528",
    doi = "10.1103/PhysRevD.92.124019",
    journal = "Phys. Rev. D",
    volume = "92",
    number = "12",
    pages = "124019",
    year = "2015"
}

@article{Bonanno:2000ep,
    author = "Bonanno, Alfio and Reuter, Martin",
    title = "{Renormalization group improved black hole space-times}",
    eprint = "hep-th/0002196",
    archivePrefix = "arXiv",
    reportNumber = "INFN-CT-03-00, MZ-TH-00-04",
    doi = "10.1103/PhysRevD.62.043008",
    journal = "Phys. Rev. D",
    volume = "62",
    pages = "043008",
    year = "2000"
}

@article{Bonanno:2006eu,
    author = "Bonanno, A. and Reuter, M.",
    title = "{Spacetime structure of an evaporating black hole in quantum gravity}",
    eprint = "hep-th/0602159",
    archivePrefix = "arXiv",
    reportNumber = "MZ-TH-06-04",
    doi = "10.1103/PhysRevD.73.083005",
    journal = "Phys. Rev. D",
    volume = "73",
    pages = "083005",
    year = "2006"
}

@book{Reuter:2019byg,
    author = "Reuter, Martin and Saueressig, Frank",
    title = "{Quantum Gravity and the Functional Renormalization Group}: {The Road towards Asymptotic Safety}",
    isbn = "978-1-107-10732-8, 978-1-108-67074-6",
    publisher = "Cambridge University Press",
    month = "1",
    year = "2019"
}

@article{Shaposhnikov:2009pv,
    author = "Shaposhnikov, Mikhail and Wetterich, Christof",
    title = "{Asymptotic safety of gravity and the Higgs boson mass}",
    eprint = "0912.0208",
    archivePrefix = "arXiv",
    primaryClass = "hep-th",
    doi = "10.1016/j.physletb.2009.12.022",
    journal = "Phys. Lett. B",
    volume = "683",
    pages = "196--200",
    year = "2010"
}

@article{Eichhorn:2017lry,
    author = "Eichhorn, Astrid and Versteegen, Fleur",
    title = "{Upper bound on the Abelian gauge coupling from asymptotic safety}",
    eprint = "1709.07252",
    archivePrefix = "arXiv",
    primaryClass = "hep-th",
    doi = "10.1007/JHEP01(2018)030",
    journal = "JHEP",
    volume = "01",
    pages = "030",
    year = "2018"
}

@article{Eichhorn:2018whv,
    author = "Eichhorn, Astrid and Held, Aaron",
    title = "{Mass difference for charged quarks from asymptotically safe quantum gravity}",
    eprint = "1803.04027",
    archivePrefix = "arXiv",
    primaryClass = "hep-th",
    doi = "10.1103/PhysRevLett.121.151302",
    journal = "Phys. Rev. Lett.",
    volume = "121",
    number = "15",
    pages = "151302",
    year = "2018"
}

@article{Eichhorn:2017ylw,
    author = "Eichhorn, Astrid and Held, Aaron",
    title = "{Top mass from asymptotic safety}",
    eprint = "1707.01107",
    archivePrefix = "arXiv",
    primaryClass = "hep-th",
    doi = "10.1016/j.physletb.2017.12.040",
    journal = "Phys. Lett. B",
    volume = "777",
    pages = "217--221",
    year = "2018"
}

@article{Dona:2013qba,
    author = "Don{\`a}, Pietro and Eichhorn, Astrid and Percacci, Roberto",
    title = "{Matter matters in asymptotically safe quantum gravity}",
    eprint = "1311.2898",
    archivePrefix = "arXiv",
    primaryClass = "hep-th",
    doi = "10.1103/PhysRevD.89.084035",
    journal = "Phys. Rev. D",
    volume = "89",
    number = "8",
    pages = "084035",
    year = "2014"
}

@article{Eichhorn:2025sux,
    author = "Eichhorn, Astrid and Gyftopoulos, Zois and Held, Aaron",
    title = "{Quark and lepton mixing in the asymptotically safe Standard Model}",
    eprint = "2507.18304",
    archivePrefix = "arXiv",
    primaryClass = "hep-ph",
    month = "7",
    year = "2025"
}

@book{Percacci:2017fkn,
    author = "Percacci, Robert",
    title = "{An Introduction to Covariant Quantum Gravity and Asymptotic Safety}",
    doi = "10.1142/10369",
    isbn = "978-981-320-717-2, 978-981-320-719-6",
    publisher = "World Scientific",
    series = "100 Years of General Relativity",
    volume = "3",
    year = "2017"
}

@article{Ashtekar:2021kfp,
    author = "Ashtekar, Abhay and Bianchi, Eugenio",
    title = "{A short review of loop quantum gravity}",
    eprint = "2104.04394",
    archivePrefix = "arXiv",
    primaryClass = "gr-qc",
    doi = "10.1088/1361-6633/abed91",
    journal = "Rept. Prog. Phys.",
    volume = "84",
    number = "4",
    pages = "042001",
    year = "2021"
}

@article{Bonanno:2019ilz,
    author = "Bonanno, Alfio and Casadio, Roberto and Platania, Alessia",
    title = "{Gravitational antiscreening in stellar interiors}",
    eprint = "1910.11393",
    archivePrefix = "arXiv",
    primaryClass = "gr-qc",
    doi = "10.1088/1475-7516/2020/01/022",
    journal = "JCAP",
    volume = "01",
    pages = "022",
    year = "2020"
}

@article{Held:2021vwd,
    author = "Held, Aaron",
    title = "{Invariant Renormalization-Group improvement}",
    eprint = "2105.11458",
    archivePrefix = "arXiv",
    primaryClass = "gr-qc",
    reportNumber = "Imperial/TP/2021/AH/04",
    month = "5",
    year = "2021"
}

@inbook{Weinberg:1980gg,
    author = "Weinberg, Steven",
    title = "{Ultraviolet divergences in quantum theories of gravitation}",
    booktitle = "{General Relativity}: {An Einstein Centenary Survey}",
    pages = "790--831",
    year = "1980"
}

@article{Wetterich:1992yh,
    author = "Wetterich, Christof",
    title = "{Exact evolution equation for the effective potential}",
    eprint = "1710.05815",
    archivePrefix = "arXiv",
    primaryClass = "hep-th",
    reportNumber = "HD-THEP-92-61",
    doi = "10.1016/0370-2693(93)90726-X",
    journal = "Phys. Lett. B",
    volume = "301",
    pages = "90--94",
    year = "1993"
}

@article{Reuter:1996cp,
    author = "Reuter, M.",
    title = "{Nonperturbative evolution equation for quantum gravity}",
    eprint = "hep-th/9605030",
    archivePrefix = "arXiv",
    reportNumber = "DESY-96-080",
    doi = "10.1103/PhysRevD.57.971",
    journal = "Phys. Rev. D",
    volume = "57",
    pages = "971--985",
    year = "1998"
}

@article{Bonanno:1998ye,
    author = "Bonanno, Alfio and Reuter, Martin",
    title = "{Quantum gravity effects near the null black hole singularity}",
    eprint = "gr-qc/9811026",
    archivePrefix = "arXiv",
    reportNumber = "INFN-CT-12-98, MZ-TH-98-45",
    doi = "10.1103/PhysRevD.60.084011",
    journal = "Phys. Rev. D",
    volume = "60",
    pages = "084011",
    year = "1999"
}

@article{Platania:2020knd,
    author = "Platania, Alessia and Wetterich, Christof",
    title = "{Non-perturbative unitarity and fictitious ghosts in quantum gravity}",
    eprint = "2009.06637",
    archivePrefix = "arXiv",
    primaryClass = "hep-th",
    doi = "10.1016/j.physletb.2020.135911",
    journal = "Phys. Lett. B",
    volume = "811",
    pages = "135911",
    year = "2020"
}

@article{Platania:2022gtt,
    author = "Platania, Alessia",
    title = "{Causality, unitarity and stability in quantum gravity: a non-perturbative perspective}",
    eprint = "2206.04072",
    archivePrefix = "arXiv",
    primaryClass = "hep-th",
    doi = "10.1007/JHEP09(2022)167",
    journal = "JHEP",
    volume = "09",
    pages = "167",
    year = "2022"
}

@article{Bonanno:2021squ,
    author = "Bonanno, Alfio and Denz, Tobias and Pawlowski, Jan M. and Reichert, Manuel",
    title = "{Reconstructing the graviton}",
    eprint = "2102.02217",
    archivePrefix = "arXiv",
    primaryClass = "hep-th",
    doi = "10.21468/SciPostPhys.12.1.001",
    journal = "SciPost Phys.",
    volume = "12",
    number = "1",
    pages = "001",
    year = "2022"
}

@article{Pastor-Gutierrez:2022nki,
    author = "Pastor-Guti\'errez, \'Alvaro and Pawlowski, Jan M. and Reichert, Manuel",
    title = "{The Asymptotically Safe Standard Model: From quantum gravity to dynamical chiral symmetry breaking}",
    eprint = "2207.09817",
    archivePrefix = "arXiv",
    primaryClass = "hep-th",
    month = "7",
    journal={SciPost Phys.},
	volume={15},
	pages={105},
	publisher={SciPost},
	doi={10.21468/SciPostPhys.15.3.105},
	url={https://scipost.org/10.21468/SciPostPhys.15.3.105},
    year = "2022"
}

@article{Knorr:2021slg,
    author = "Knorr, Benjamin",
    title = "{The derivative expansion in asymptotically safe quantum gravity: general setup and quartic order}",
    eprint = "2104.11336",
    archivePrefix = "arXiv",
    primaryClass = "hep-th",
    doi = "10.21468/SciPostPhysCore.4.3.020",
    journal = "SciPost Phys. Core",
    volume = "4",
    pages = "020",
    year = "2021"
}

@article{Gies:2016con,
    author = "Gies, Holger and Knorr, Benjamin and Lippoldt, Stefan and Saueressig, Frank",
    title = "{Gravitational Two-Loop Counterterm Is Asymptotically Safe}",
    eprint = "1601.01800",
    archivePrefix = "arXiv",
    primaryClass = "hep-th",
    doi = "10.1103/PhysRevLett.116.211302",
    journal = "Phys. Rev. Lett.",
    volume = "116",
    number = "21",
    pages = "211302",
    year = "2016"
}

@article{Knorr:2021niv,
    author = "Knorr, Benjamin and Schiffer, Marc",
    title = "{Non-Perturbative Propagators in Quantum Gravity}",
    eprint = "2105.04566",
    archivePrefix = "arXiv",
    primaryClass = "hep-th",
    doi = "10.3390/universe7070216",
    journal = "Universe",
    volume = "7",
    number = "7",
    pages = "216",
    year = "2021"
}

@article{Knorr:2019atm,
    author = "Knorr, Benjamin and Ripken, Chris and Saueressig, Frank",
    title = "{Form Factors in Asymptotic Safety: conceptual ideas and computational toolbox}",
    eprint = "1907.02903",
    archivePrefix = "arXiv",
    primaryClass = "hep-th",
    doi = "10.1088/1361-6382/ab4a53",
    journal = "Class. Quant. Grav.",
    volume = "36",
    number = "23",
    pages = "234001",
    year = "2019"
}

@article{Bonanno:2019rsq,
    author = "Bonanno, Alfio and Silveravalle, Samuele",
    title = "{Characterizing black hole metrics in quadratic gravity}",
    eprint = "1903.08759",
    archivePrefix = "arXiv",
    primaryClass = "gr-qc",
    doi = "10.1103/PhysRevD.99.101501",
    journal = "Phys. Rev. D",
    volume = "99",
    number = "10",
    pages = "101501",
    year = "2019"
}

@phdthesis{Silveravalle:2023lnl,
    author = "Silveravalle, Samuele Marco",
    title = "{Isolated Objects in Quadratic Gravity: From Action Principles to Observations}",
    doi = "10.1007/978-3-031-48994-5",
    school = "Trento U.",
    year = "2024"
}

@article{Goldstein:2017rxn,
    author = "Goldstein, Kevin and Mashiyane, James Junior",
    title = "{Ineffective Higher Derivative Black Hole Hair}",
    eprint = "1703.02803",
    archivePrefix = "arXiv",
    primaryClass = "hep-th",
    doi = "10.1103/PhysRevD.97.024015",
    journal = "Phys. Rev. D",
    volume = "97",
    number = "2",
    pages = "024015",
    year = "2018"
}

@article{Podolsky:2018pfe,
    author = "Podolsky, Jiri and Svarc, Robert and Pravda, Vojtech and Pravdova, Alena",
    title = "{Explicit black hole solutions in higher-derivative gravity}",
    eprint = "1806.08209",
    archivePrefix = "arXiv",
    primaryClass = "gr-qc",
    doi = "10.1103/PhysRevD.98.021502",
    journal = "Phys. Rev. D",
    volume = "98",
    number = "2",
    pages = "021502",
    year = "2018"
}

@article{Podolsky:2019gro,
    author = "Podolsk\'y, Jiri and \v{S}varc, Robert and Pravda, Vojtech and Pravdova, Alena",
    title = "{Black holes and other exact spherical solutions in Quadratic Gravity}",
    eprint = "1907.00046",
    archivePrefix = "arXiv",
    primaryClass = "gr-qc",
    doi = "10.1103/PhysRevD.101.024027",
    journal = "Phys. Rev. D",
    volume = "101",
    number = "2",
    pages = "024027",
    year = "2020"
}

@article{Bonanno:2016dyv,
    author = "Bonanno, Alfio and Koch, Benjamin and Platania, Alessia",
    title = "{Cosmic Censorship in Quantum Einstein Gravity}",
    eprint = "1610.05299",
    archivePrefix = "arXiv",
    primaryClass = "gr-qc",
    doi = "10.1088/1361-6382/aa6788",
    journal = "Class. Quant. Grav.",
    volume = "34",
    number = "9",
    pages = "095012",
    year = "2017"
}

@article{Bonanno:2017kta,
    author = "Bonanno, Alfio and Koch, Benjamin and Platania, Alessia",
    title = "{Asymptotically Safe gravitational collapse: Kuroda-Papapetrou RG-improved model}",
    doi = "10.22323/1.292.0058",
    journal = "PoS",
    volume = "CORFU2016",
    pages = "058",
    year = "2017"
}

@article{Bonanno:2017zen,
    author = "Bonanno, Alfio and Koch, Benjamin and Platania, Alessia",
    title = "{Gravitational collapse in Quantum Einstein Gravity}",
    eprint = "1710.10845",
    archivePrefix = "arXiv",
    primaryClass = "gr-qc",
    doi = "10.1007/s10701-018-0195-7",
    journal = "Found. Phys.",
    volume = "48",
    number = "10",
    pages = "1393--1406",
    year = "2018"
}

@article{Torres:2017ygl,
    author = "Torres, Ramon",
    title = "{Nonsingular black holes, the cosmological constant, and asymptotic safety}",
    eprint = "1703.09997",
    archivePrefix = "arXiv",
    primaryClass = "gr-qc",
    doi = "10.1103/PhysRevD.95.124004",
    journal = "Phys. Rev. D",
    volume = "95",
    number = "12",
    pages = "124004",
    year = "2017"
}

@article{Adeifeoba:2018ydh,
    author = "Adeifeoba, Ademola and Eichhorn, Astrid and Platania, Alessia",
    title = "{Towards conditions for black-hole singularity-resolution in asymptotically safe quantum gravity}",
    eprint = "1808.03472",
    archivePrefix = "arXiv",
    primaryClass = "gr-qc",
    doi = "10.1088/1361-6382/aae6ef",
    journal = "Class. Quant. Grav.",
    volume = "35",
    number = "22",
    pages = "225007",
    year = "2018"
}

@article{Pawlowski:2018swz,
    author = "Pawlowski, Jan M. and Stock, Dennis",
    title = "{Quantum-improved Schwarzschild-(A)dS and Kerr-(A)dS spacetimes}",
    eprint = "1807.10512",
    archivePrefix = "arXiv",
    primaryClass = "hep-th",
    doi = "10.1103/PhysRevD.98.106008",
    journal = "Phys. Rev. D",
    volume = "98",
    number = "10",
    pages = "106008",
    year = "2018"
}

@article{Platania:2019kyx,
    author = "Platania, Alessia",
    title = "{Dynamical renormalization of black-hole spacetimes}",
    eprint = "1903.10411",
    archivePrefix = "arXiv",
    primaryClass = "gr-qc",
    doi = "10.1140/epjc/s10052-019-6990-2",
    journal = "Eur. Phys. J. C",
    volume = "79",
    number = "6",
    pages = "470",
    year = "2019"
}

@article{Baldazzi:2021ydj,
    author = "Baldazzi, Alessio and Zinati, Riccardo Ben Al{\`\i} and Falls, Kevin",
    title = "{Essential renormalisation group}",
    eprint = "2105.11482",
    archivePrefix = "arXiv",
    primaryClass = "hep-th",
    doi = "10.21468/SciPostPhys.13.4.085",
    journal = "SciPost Phys.",
    volume = "13",
    number = "4",
    pages = "085",
    year = "2022"
}

@article{Baldazzi:2021orb,
    author = "Baldazzi, Alessio and Falls, Kevin",
    title = "{Essential Quantum Einstein Gravity}",
    eprint = "2107.00671",
    archivePrefix = "arXiv",
    primaryClass = "hep-th",
    doi = "10.3390/universe7080294",
    journal = "Universe",
    volume = "7",
    number = "8",
    pages = "294",
    year = "2021"
}

@article{Basile:2021krr,
    author = "Basile, Ivano and Platania, Alessia",
    title = "{Asymptotic Safety: Swampland or Wonderland?}",
    eprint = "2107.06897",
    archivePrefix = "arXiv",
    primaryClass = "hep-th",
    doi = "10.3390/universe7100389",
    journal = "Universe",
    volume = "7",
    number = "10",
    pages = "389",
    year = "2021"
}

@article{Emoto:2005te,
      author         = "Emoto, Hiroki",
      title          = "{Asymptotic safety of quantum gravity and improved
                        spacetime of black hole singularity by cutoff
                        identification}",
      year           = "2005",
      eprint         = "hep-th/0511075",
      archivePrefix  = "arXiv",
      primaryClass   = "hep-th",
      SLACcitation   = "%%CITATION = HEP-TH/0511075;%%"
}

@inproceedings{Reuter:2006rg,
      author         = "Reuter, M. and Tuiran, E.",
      title          = "{Quantum Gravity Effects in Rotating Black Holes}",
      booktitle      = "{Recent developments in theoretical and experimental
                        general relativity, gravitation and relativistic field
                        theories. Proceedings, 11th Marcel Grossmann Meeting,
                        MG11, Berlin, Germany, July 23-29, 2006. Pt. A-C}",
      year           = "2006",
      pages          = "2608-2610",
      doi            = "10.1142/9789812834300_0473",
      eprint         = "hep-th/0612037",
      archivePrefix  = "arXiv",
      primaryClass   = "hep-th",
      reportNumber   = "MZ-TH-06-24",
      SLACcitation   = "%%CITATION = HEP-TH/0612037;%%"
}

@article{Falls:2010he,
      author         = "Falls, Kevin and Litim, Daniel F. and Raghuraman, Aarti",
      title          = "{Black Holes and Asymptotically Safe Gravity}",
      journal        = "Int. J. Mod. Phys.",
      volume         = "A27",
      year           = "2012",
      pages          = "1250019",
      doi            = "10.1142/S0217751X12500194",
      eprint         = "1002.0260",
      archivePrefix  = "arXiv",
      primaryClass   = "hep-th",
      SLACcitation   = "%%CITATION = ARXIV:1002.0260;%%"
}

@article{Cai:2010zh,
      author         = "Cai, Yi-Fu and Easson, Damien A.",
      title          = "{Black holes in an asymptotically safe gravity theory
                        with higher derivatives}",
      journal        = "JCAP",
      volume         = "1009",
      year           = "2010",
      pages          = "002",
      doi            = "10.1088/1475-7516/2010/09/002",
      eprint         = "1007.1317",
      archivePrefix  = "arXiv",
      primaryClass   = "hep-th",
      reportNumber   = "IPMU-10-0103, NSF-KITP-10-085",
      SLACcitation   = "%%CITATION = ARXIV:1007.1317;%%"
}

@article{Fayos:2011zza,
      author         = "Fayos, F. and Torres, R.",
      title          = "{A quantum improvement to the gravitational collapse of
                        radiating stars}",
      journal        = "Class. Quant. Grav.",
      volume         = "28",
      year           = "2011",
      pages          = "105004",
      doi            = "10.1088/0264-9381/28/10/105004",
      SLACcitation   = "%%CITATION = CQGRD,28,105004;%%"
}

@article{Falls:2012nd,
      author         = "Falls, Kevin and Litim, Daniel F.",
      title          = "{Black hole thermodynamics under the microscope}",
      journal        = "Phys. Rev.",
      volume         = "D89",
      year           = "2014",
      pages          = "084002",
      doi            = "10.1103/PhysRevD.89.084002",
      eprint         = "1212.1821",
      archivePrefix  = "arXiv",
      primaryClass   = "gr-qc",
      SLACcitation   = "%%CITATION = ARXIV:1212.1821;%%"
}

@article{Koch:2013owa,
      author         = "Koch, Benjamin and Saueressig, Frank",
      title          = "{Structural aspects of asymptotically safe black holes}",
      journal        = "Class. Quant. Grav.",
      volume         = "31",
      year           = "2014",
      pages          = "015006",
      doi            = "10.1088/0264-9381/31/1/015006",
      eprint         = "1306.1546",
      archivePrefix  = "arXiv",
      primaryClass   = "hep-th",
      SLACcitation   = "%%CITATION = ARXIV:1306.1546;%%"
}

@article{Held:2019xde,
      author         = "Held, Aaron and Gold, Roman and Eichhorn, Astrid",
      title          = "{Asymptotic safety casts its shadow}",
      journal        = "JCAP",
      volume         = "1906",
      year           = "2019",
      number         = "06",
      pages          = "029",
      doi            = "10.1088/1475-7516/2019/06/029",
      eprint         = "1904.07133",
      archivePrefix  = "arXiv",
      primaryClass   = "gr-qc",
      SLACcitation   = "%%CITATION = ARXIV:1904.07133;%%"
}

@article{Borissova:2020knn,
	author = "Borissova, Johanna N. and Eichhorn, Astrid",
	title = "{Towards black-hole singularity-resolution in the Lorentzian gravitational path integral}",
	eprint = "2012.08570",
	archivePrefix = "arXiv",
	primaryClass = "gr-qc",
	doi = "10.3390/universe7030048",
	journal = "Universe",
	volume = "7",
	number = "3",
	pages = "48",
	year = "2021"
}

@article{Knorr:2024yiu,
    author = "Knorr, Benjamin and Platania, Alessia",
    title = "{Unearthing the intersections: positivity bounds, weak gravity conjecture, and asymptotic safety landscapes from photon-graviton flows}",
    eprint = "2405.08860",
    archivePrefix = "arXiv",
    primaryClass = "hep-th",
    reportNumber = "NORDITA 2024-014",
    doi = "10.1007/JHEP03(2025)003",
    journal = "JHEP",
    volume = "03",
    pages = "003",
    year = "2025"
}

@article{AparicioResco:2016xcm,
    author = "Aparicio Resco, Miguel and de la Cruz-Dombriz, {\'A}lvaro and Llanes Estrada, Felipe J. and Zapatero Castrillo, V{\'\i}ctor",
    title = "{On neutron stars in $f(R)$ theories: Small radii, large masses and large energy emitted in a merger}",
    eprint = "1602.03880",
    archivePrefix = "arXiv",
    primaryClass = "gr-qc",
    doi = "10.1016/j.dark.2016.07.001",
    journal = "Phys. Dark Univ.",
    volume = "13",
    pages = "147--161",
    year = "2016"
}

@article{Astashenok:2017dpo,
    author = "Astashenok, Artyom V. and Odintsov, Sergei D. and de la Cruz-Dombriz, Alvaro",
    title = "{The realistic models of relativistic stars in $f(R) = R + \alpha R^2$ gravity}",
    eprint = "1704.08311",
    archivePrefix = "arXiv",
    primaryClass = "gr-qc",
    doi = "10.1088/1361-6382/aa8971",
    journal = "Class. Quant. Grav.",
    volume = "34",
    number = "20",
    pages = "205008",
    year = "2017"
}

@article{Sbisa:2019mae,
    author = "Sbis{\`a}, Fulvio and Baqui, Pedro O. and Miranda, Tays and Jor{\'a}s, Sergio E. and Piattella, Oliver F.",
    title = "{Neutron star masses in $R^{2}$-gravity}",
    eprint = "1907.08714",
    archivePrefix = "arXiv",
    primaryClass = "gr-qc",
    doi = "10.1016/j.dark.2019.100411",
    journal = "Phys. Dark Univ.",
    volume = "27",
    pages = "100411",
    year = "2020"
}

@article{Bonanno:2021zoy,
    author = "Bonanno, Alfio and Silveravalle, Samuele",
    title = "{The gravitational field of a star in quadratic gravity}",
    eprint = "2106.00558",
    archivePrefix = "arXiv",
    primaryClass = "gr-qc",
    doi = "10.1088/1475-7516/2021/08/050",
    journal = "JCAP",
    volume = "08",
    pages = "050",
    year = "2021"
}

@article{Antoniou:2024jku,
    author = "Antoniou, Georgios and Gualtieri, Leonardo and Pani, Paolo",
    title = "{Gravitational quasinormal modes of black holes in quadratic gravity}",
    eprint = "2412.15037",
    archivePrefix = "arXiv",
    primaryClass = "gr-qc",
    doi = "10.1103/PhysRevD.111.064059",
    journal = "Phys. Rev. D",
    volume = "111",
    number = "6",
    pages = "064059",
    year = "2025"
}

@article{East:2023nsk,
    author = "East, William E. and Siemonsen, Nils",
    title = "{Instability and backreaction of massive spin-2 fields around black holes}",
    eprint = "2309.05096",
    archivePrefix = "arXiv",
    primaryClass = "gr-qc",
    doi = "10.1103/PhysRevD.108.124048",
    journal = "Phys. Rev. D",
    volume = "108",
    number = "12",
    pages = "124048",
    year = "2023"
}

@article{Holdom:2022zzo,
    author = "Holdom, Bob",
    title = "{2-2-holes simplified}",
    eprint = "2202.08442",
    archivePrefix = "arXiv",
    primaryClass = "gr-qc",
    doi = "10.1016/j.physletb.2022.137142",
    journal = "Phys. Lett. B",
    volume = "830",
    pages = "137142",
    year = "2022"
}

@article{Casadio:2010fw,
      author         = "Casadio, Roberto and Hsu, Stephen D. H. and Mirza,
                        Behrouz",
      title          = "{Asymptotic Safety, Singularities, and Gravitational
                        Collapse}",
      journal        = "Phys. Lett.",
      volume         = "B695",
      year           = "2011",
      pages          = "317-319",
      doi            = "10.1016/j.physletb.2010.10.060",
      eprint         = "1008.2768",
      archivePrefix  = "arXiv",
      primaryClass   = "gr-qc",
      SLACcitation   = "%%CITATION = ARXIV:1008.2768;%%"
}

@article{Chen:2023pcv,
    author = "Chen, Chiang-Mei and Chen, Yi and Ishibashi, Akihiro and Ohta, Nobuyoshi",
    title = "{Phase structure of quantum improved Schwarzschild-(Anti)de Sitter black holes}",
    eprint = "2303.04304",
    archivePrefix = "arXiv",
    primaryClass = "hep-th",
    doi = "10.1088/1361-6382/acfc91",
    journal = "Class. Quant. Grav.",
    volume = "40",
    number = "21",
    pages = "215007",
    year = "2023"
}

@article{Bonanno:2023rzk,
    author = "Bonanno, Alfio and Malafarina, Daniele and Panassiti, Antonio",
    title = "{Dust Collapse in Asymptotic Safety: A Path to Regular Black Holes}",
    eprint = "2308.10890",
    archivePrefix = "arXiv",
    primaryClass = "gr-qc",
    doi = "10.1103/PhysRevLett.132.031401",
    journal = "Phys. Rev. Lett.",
    volume = "132",
    number = "3",
    pages = "031401",
    year = "2024"
}

@article{Horowitz:2023xyl,
    author = "Horowitz, Gary T. and Kolanowski, Maciej and Remmen, Grant N. and Santos, Jorge E.",
    title = "{Extremal Kerr Black Holes as Amplifiers of New Physics}",
    eprint = "2303.07358",
    archivePrefix = "arXiv",
    primaryClass = "hep-th",
    doi = "10.1103/PhysRevLett.131.091402",
    journal = "Phys. Rev. Lett.",
    volume = "131",
    number = "9",
    pages = "091402",
    year = "2023"
}

@article{Baldazzi:2023pep,
    author = "Baldazzi, Alessio and Falls, Kevin and Kluth, Yannick and Knorr, Benjamin",
    title = "{Robustness of the derivative expansion in Asymptotic Safety}",
    eprint = "2312.03831",
    archivePrefix = "arXiv",
    primaryClass = "hep-th",
    reportNumber = "NORDITA 2023-075",
    month = "12",
    year = "2023"
}

@article{Goroff:1985sz,
    author = "Goroff, Marc H. and Sagnotti, Augusto",
    title = "{QUANTUM GRAVITY AT TWO LOOPS}",
    reportNumber = "CALT-68-1263, UCB-PTH-85/18, LBL-19512",
    doi = "10.1016/0370-2693(85)91470-4",
    journal = "Phys. Lett. B",
    volume = "160",
    pages = "81--86",
    year = "1985"
}

@article{Goroff:1985th,
    author = "Goroff, Marc H. and Sagnotti, Augusto",
    title = "{The Ultraviolet Behavior of Einstein Gravity}",
    reportNumber = "CALT-68-1289, LBL-19995, UCB-PTH-85-34",
    doi = "10.1016/0550-3213(86)90193-8",
    journal = "Nucl. Phys. B",
    volume = "266",
    pages = "709--736",
    year = "1986"
}

@article{vandeVen:1991gw,
    author = "van de Ven, A. E. M.",
    title = "{Two loop quantum gravity}",
    reportNumber = "DESY-91-115, ITP-SB-91-52",
    doi = "10.1016/0550-3213(92)90011-Y",
    journal = "Nucl. Phys. B",
    volume = "378",
    pages = "309--366",
    year = "1992"
}

@article{Borissova:2023kzq,
    author = "Borissova, Johanna N.",
    title = "{Suppression of spacetime singularities in quantum gravity}",
    eprint = "2309.05695",
    archivePrefix = "arXiv",
    primaryClass = "gr-qc",
    doi = "10.1088/1361-6382/ad46c0",
    journal = "Class. Quant. Grav.",
    volume = "41",
    number = "12",
    pages = "127002",
    year = "2024"
}

@article{Gies:2015tca,
    title = {{Generalized Parametrization Dependence in Quantum Gravity}},
    year = {2015},
    journal = {Phys. Rev.},
    author = {Gies, Holger and Knorr, Benjamin and Lippoldt, Stefan},
    number = {8},
    pages = {84020},
    volume = {D92},
    doi = {10.1103/PhysRevD.92.084020},
    arxivId = {hep-th/1507.08859}
}

@article{DelPorro:2025fiu,
    author = "Del Porro, Francesco and Ferrarin, Francesco and Platania, Alessia",
    title = "{Impact of quantum gravity on the UV sensitivity of extremal black holes}",
    eprint = "2509.07058",
    archivePrefix = "arXiv",
    primaryClass = "hep-th",
    month = "9",
    year = "2025"
}

@article{Giacchini:2025mlv,
    author = "Giacchini, Breno L. and Kol{\'a}{\v{r}}, Ivan",
    title = "{Neglected solutions in quadratic gravity}",
    eprint = "2509.07317",
    archivePrefix = "arXiv",
    primaryClass = "gr-qc",
    month = "9",
    year = "2025"
}

@article{Souma:1999at,
    title = {{Non-Trivial Ultraviolet Fixed Point in Quantum Gravity}},
    year = {1999},
    journal = {Progress of Theoretical Physics},
    author = {Souma, W},
    month = {7},
    pages = {181--195},
    volume = {102},
    doi = {10.1143/PTP.102.181}
}

\end{document}